\documentclass[preprint,review,12pt,authoryear,3p]{elsarticle}

\usepackage[T1]{fontenc}

\usepackage{amssymb}
\usepackage{amsmath}
\usepackage{lineno}
\usepackage{array}
\usepackage{multirow}
\usepackage{comment}

\renewcommand{\vec}[1]{\mathbf{#1}}
\newcommand{\tens}[1]{\boldsymbol{\underline{\underline{#1}}}}
\newcommand{\nb}[1]{\vec{{n}}}
\newcommand{\nhat}[1]{\vec{\hat{n}}}
\newcommand{\that}[1]{\vec{\hat{t}}_1}
\newcommand{\thati}[1]{\vec{\hat{t}}_2}

\journal{ArXiv}

\begin{document}

\begin{frontmatter}



\title{Generalisation of the Navier-slip boundary condition to arbitrary directions: Application to 3D oblique geodynamic simulations}

\author[label1]{Anthony Jourdon} 
\author[label2]{Dave A. May}
\author[label2,label1]{Alice-Agnes Gabriel}

\affiliation[label1]{organization={Department of Earth and Environmental Sciences, Ludwig-Maximillians-Universität München},
            city={Munich},
            country={Germany}}
\affiliation[label2]{organization={Institute of Geophysics and Planetary Physics, Scripps Institution of Oceanography, UC San Diego},
            city={La Jolla},
            state={CA},
            country={USA}}

\begin{abstract}
Although boundary conditions are mandatory to solve partial differential equations, they also represent a transfer of information between the domain being modelled and its surroundings.
In the case of isolated or closed systems, these can be formulated using free- or no-slip conditions. 
However, for other types of system, the information transferred through the boundaries is essential to the dynamics of the system and can have a first order impact on its evolution.

This work addresses regional geodynamic modelling, which simulates the evolution of an Earth's piece over millions of years by solving non-linear Stokes flow. 
In this open system, we introduce a new approach to impose oblique boundary conditions.
These new conditions generalise the Navier-slip boundary conditions to arbitrary directions in three dimensions. 
The method requires defining both slip and stress constraints.
The stress constraint is imposed utilising a coordinate transformation to redefine the stress tensor along the boundaries according to the arbitrary direction chosen while for the slip constraint we utilise Nitsche's approach in the context of the finite element method, resulting in a symmetrised and penalised weak form. 
We validate our approach through a series of numerical experiments of increasing complexity, starting with 2D and 3D linear models. 
Then, we apply those boundary conditions to a 3D non-linear geodynamic model of oblique extension that we compare with a standard model utilising Dirichlet boundary conditions.
Our results show that using Dirichlet boundary conditions, which requires to provide arbitrary velocity functions along the boundaries, strongly influences the evolution of the system and generates artefacts near and along the boundaries.
In comparison, the model using the generalised Navier-slip boundary conditions behaves closely to a model with an unbounded domain, providing a physically interpretable solution near and along the boundaries.
Finally, we show that the numerical solve of the system is not affected by these boundary conditions when using Krylov methods preconditioned with multigrid.
The method presented in this work offers a more accurate and physically meaningful approach to impose oblique boundary conditions to an open system which is a first order improvement for regional geodynamic simulations.
\end{abstract}

\begin{highlights}
\item Generalisation of Navier-slip boundary conditions to arbitrary directions. 
\item Impose oblique boundary conditions by constraining velocity direction, not magnitude.
\item Simulation of regional-scale long-term oblique geodynamic systems in 3D.
\item Utilisation of Nitsche's method to impose slip constraint in arbitrary directions.
\end{highlights}


\end{frontmatter}

\section{Introduction}
\label{sec:intro}
\subsection{Geodynamic modelling} 
\label{sub:geodynamic_modelling}

Long term geodynamics modelling studies the motion of Earth's rocks over millions of years. 
To physically model large space and time scales motions the Earth's interior is treated as non-linear highly viscous fluids that are accounted for in the Stokes or Navier-Stokes equations.
Long term geodynamics can be modelled through two approaches. One approach consists of modelling the entire planet, the global models \citep[e.g.][]{Tackley2008,Kronbichler2012,Heister2017} while the other approach focuses only on a region, the regional models \citep[e.g.][]{Gerya2007,Popov2008,May2015,Kaus2016}.
In this work we focus on regional models.

Regional models present the advantage of being relatively small with respect to the Earth ($\sim10^{3}$ km in each spatial direction) allowing high spatial resolution models ($\sim10^{3}$ m in each spatial direction).
However, because they represent a region embedded in a non-modelled global domain they require boundary conditions in the three spatial dimensions. 
These boundary conditions are not only a mandatory mathematical formulation to uniquely solve a system of partial differential equations, they also represent a transfer of physical information. 
Obviously, depending on the transferred information the mechanical response of the modelled domain will differ. 

On Earth, the internal dynamics of the planet notably expresses through the motion  of tectonic plates leading to the formation of oceanic ridges, mountain belts and subduction zones. 
These particular places are called tectonic plates boundaries and represent the main expression of Earth's activity concentrating earthquakes, mineral resources and human economic activity.
The first and foremost observation is that the deformation in tectonic plates boundaries is three-dimensional, i.e. non-cylindrical, oblique \citep{Bird2003,Brune2018}.

\subsection{Motivation and challenges} 
\label{sub:motivation_challenges}

In models the obliquity is mostly applied through initial conditions i.e., oblique weak zones together with free-slip boundary conditions \citep[e.g.][]{Brune2012,Ammann2017,LePourhiet2017,Duclaux2020}. 

\begin{figure}[h!]
    \centering
    \includegraphics[scale=1]{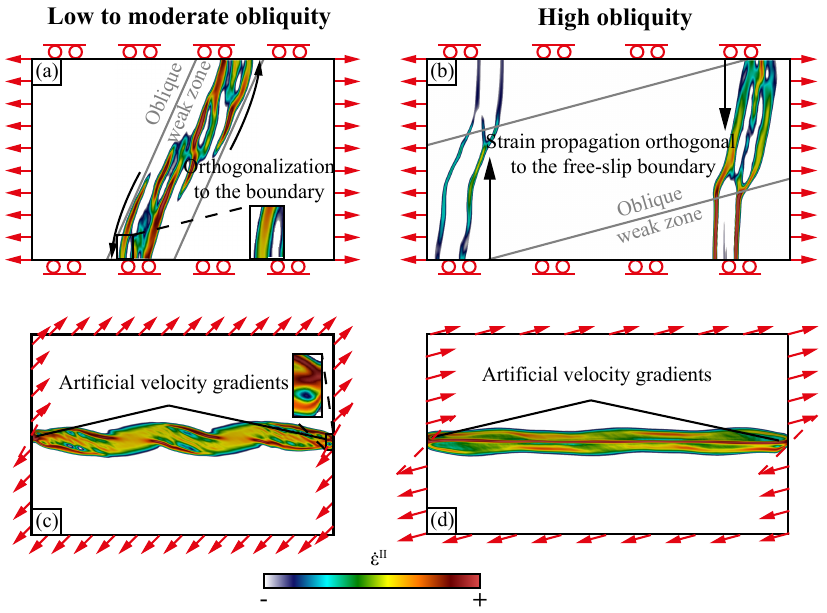}
    \caption{Map views of 3D geodynamic numerical models modelling oblique stretching.
(a) Low to moderate obliquity, (b) high obliquity imposed through initial conditions (weak zones) coupled with free-slip boundary condition.
In both cases, the free-slip boundary condition forces the deformation to orthogonalize while approaching the boundary. 
High obliquity model shows a total obliteration of the obliquity of the initial weak zone.
(c) Low to moderate obliquity, (d) high obliquity imposed through Dirichlet boundary conditions.
In both cases the transition of the velocity vectors from one direction to the opposite is arbitrary and enforces artificial gradients along the boundaries.}
    \label{fig:obliquity-methods}
\end{figure}

However, the free-slip boundary condition represents an infinitely resistant material in the direction normal to the boundary and a null friction material in the direction tangential to the boundary.
The problem with that condition is that it enforces cylindrical behaviours in the vicinity of the boundary, limiting the obliquity of the whole system or forcing to consider very large domains to avoid a too strong influence of the boundary conditions.
Figure~\ref{fig:obliquity-methods}a~\&~\ref{fig:obliquity-methods}b illustrates such issues.
In case of low to moderate obliquity (Figure~\ref{fig:obliquity-methods}a) the influence of the free-slip boundary condition is confined to the vicinity of the boundaries.
However, in case of high obliquity (Figure~\ref{fig:obliquity-methods}b) the impact of the free-slip condition influences the evolution of the whole system by enforcing a cylindrical behaviour and a strain propagation orthogonal to the boundaries.
The latter case illustrates the strong limitations caused by the free-slip boundary condition.

Therefore, to work around this problem, some studies proposed to impose obliquity through boundary conditions \citep[e.g.][]{Brune2012,Brune2014a,Heine2014,LePourhiet2018a,Jourdon2020a,Jourdon2021}.
Until now, the only approach to impose oblique boundary conditions involves strong Dirichlet imposition i.e., directly setting a value to the velocity (or displacement) along the boundary. 
This approach succeeds to simulate highly oblique systems \citep{Jourdon2021}.
Nevertheless, the problem is that requires to provide values to the velocity vectors everywhere along the boundaries and particularly in regions where the vectors are pointing in opposite directions leading to artificial and arbitrarily imposed velocity gradients in the tangential direction of the boundary (Figure~\ref{fig:obliquity-methods}c~\&~\ref{fig:obliquity-methods}d).
Such boundary effects can then influence the strain localization and produce non physical results.
Moreover, strong Dirichlet boundary conditions remains simple for Eulerian domains whose boundaries are aligned with the coordinate system, but for arbitrary Eulerian or Lagrangian meshes, being able to directly set a value to the velocity vector components is, in general, not trivial.

Thus, there is still a necessity to find a method to impose obliquity in geodynamic models without imposing such strong limitations to the modelling objectives.

\subsection{Previous work} 
\label{sub:previous_work}

The objective is to impose oblique boundary conditions i.e., the velocity direction but without constraining the magnitude of the velocity vectors to avoid arbitrary choices leading to artificial velocity gradients along the boundaries. 
Therefore, we seek to impose a slip-type boundary condition.

In the finite element method several approaches can be considered to apply a slip-type boundary condition.
The Lagrange multiplier method was used to impose free-slip boundary conditions on curved surfaces \citep{Verfurth1987}.
This approach introduces a new variable and requires particular choices of finite elements function space \citep{Verfurth1987} or the addition of a penalty term in the weak form resulting in a stabilized formulation \citep{Verfurth1991,Urquiza2014}.
Nonetheless, \cite{Stenberg1995} showed that the stabilized Lagrange multiplier method was closely related to Nitsche's method \citep{Nitsche1971} to weakly impose Dirichlet boundary conditions.
Nitsche's method was firstly developed to weakly impose Dirichlet boundary conditions in Poisson-type equations \citep{Nitsche1971} before being extended to Stokes and Navier-Stokes equations \citep{Stenberg1995}.
Then, \cite{Freund1995} adapted the method to weakly impose free-slip boundary conditions.
The method is based on the symmetrization of the Neumann stress boundary term and the penalization of the slip condition in the boundary integral \citep{Nitsche1971,Stenberg1995}. 

\subsection{Present work} 
\label{sub:present_work}

In this work, we propose a formulation in terms of slip-type conditions to impose the direction of the velocity vector and compute its magnitude based on stress.
This formulation is based on the finite element method and is a generalisation of Nitsche's method to arbitrary directions to apply Navier-slip boundary conditions.
Firstly we provide the variational form in the context of the incompressible Stokes equation.
Secondly we show simple 2D and 3D numerical simulations employing this method to model a rotated Couette flow and an oblique stretching.
Thirdly we provide two geodynamic models of oblique extension. The first model uses classical Dirichlet boundary conditions while the second model employs the generalised Navier-slip boundary conditions.
Finally, we compare these results both in term of impact on the geodynamic evolution and numerical efficiency. 

\section{Governing equations}
\label{sec:gov-eqs}
\subsection{Conservation of mass and momentum}
\label{sub:conservation}

In regional geodynamics simulations, a common physical model is to consider non-inertial incompressible flow. In this case the conservation of momentum in an open bounded domain $\Omega \in \mathbb R^3$ with boundary $\partial \Omega$ is thus:
\begin{align}
	\nabla \cdot \tens{\tau}(\vec u, p) - \nabla p + \rho \vec g &= \vec 0
	\label{eq:momentum}\\
	-\nabla \cdot \vec u &= 0,
	\label{eq:div}
\end{align}
where $\vec u$ is the velocity vector, $p$ is the pressure, $\rho$ the material density, $\vec g$ the gravity acceleration vector, 
\begin{equation}
	\tens{\tau}(\vec u, p) = 2 \eta(\vec u, p) \tens{\varepsilon}(\vec u)
\label{eq:tau}
\end{equation}
is the deviatoric stress tensor, $\eta$ the velocity and pressure dependent effective viscosity and
\begin{equation}
	\tens{\varepsilon}(\vec u) = \frac{1}{2} \left( \nabla \vec u + \nabla \vec u^T \right)
\label{eq:sr}
\end{equation}
the strain rate tensor.
The total stress is given by
\begin{equation}
	\tens{\sigma}(\vec u,\, p) = \tens{\tau}(\vec u, p) - p\tens{I},
\end{equation}
where $\tens I \in \mathbb R^3$ is the identity tensor.

\subsection{Boundary conditions}
\label{sub:BCs}

We will denote the outward pointing unit normal vector to $\partial \Omega$ by $\nb{}$.
To solve equations~\eqref{eq:momentum}~\&~\eqref{eq:div}, 
the traditional modelling approach considers the following boundary conditions
\begin{align}
	u_i &= \bar{u}_i, \quad i = 1, 2, 3 &\forall \vec x \in \Gamma_D 
	\label{eq:dirichlet}\\
	\tens{\sigma} \nb{} &= \vec{\bar{T}} &\forall \vec x \in \Gamma_N
	\label{eq:neumann}\\
	\vec u \cdot \nb{} &= \bar{g} &\forall \vec x \in \Gamma_S \\
	\tens{\tau} \nb{} &= \vec{\bar{T}} &\forall \vec x \in \Gamma_S
\end{align}
where $\vec x$ denotes the coordinates vector,
$\Gamma_D$ defines the segment where Dirichlet constraints are imposed, 
$\Gamma_N$ defines the segment where Neumann constraints are imposed
and $\Gamma_S$ defines the segment along which Navier-slip constraints are applied.

In this work, we generalise the notion of the Navier-slip conditions.
Our generalisation enables constraining the normal component of the 
velocity and shear stress across a plane defined arbitrarily with respect to the
boundary $\partial \Omega$.
We denote the unit vector normal to this arbitrary plane via $\nhat{}$.
The two tangents spanning the plane with normal $\nhat{}$ are denoted by
$\that{}$ and $\thati{}$ respectively. 
The three unit vectors $\nhat{}, \that{}, \thati{}$ define a new coordinate system satisfying
\begin{equation*}
	\begin{matrix}
		\nhat{}  \times \that{}  = \thati{}, & &
		\thati{} \times \nhat{}  = \that{}, & &
		\that{}  \times \thati{} = \nhat{}.
	\end{matrix}
\end{equation*}
The generalised Navier-slip boundary conditions express the velocity and traction constraints in terms of that newly defined coordinates system.
The velocity constraint is thus given by:
\begin{align}
	\vec u \cdot \nhat{} &= \bar{g} &\forall \vec x \in \Gamma_S . 
	\label{eq:u-slip-a}
\end{align}

To define the stress constraints we first define the rotation matrix $\tens{\Lambda}$ as
\begin{equation}
	\tens{\Lambda} :=
	\begin{bmatrix}
		\nhat{} & \that{} & \thati{}
	\end{bmatrix}
	=
	\begin{bmatrix}
		\vec{\Lambda}_0 & \vec{\Lambda}_1 & \vec{\Lambda}_2 
	\end{bmatrix}
	.
\label{eq:lambda}
\end{equation}
We also introduce $\tens{\mathcal{H}}$ a $3 \times 3$ symmetric tensor with entries $\mathcal{H}_{ij}$ given by either $0$ or $1$.
The purpose of this tensor is to discriminate which stress tensor components are constrained ($\mathcal{H}_{ij} = 1$) and which are not ($\mathcal{H}_{ij} = 0$).
We also denote by $( \odot )$ the point-wise product between tensors such that $ \tens{A} \odot \tens{B} = A_{ij} B_{ij} \quad \forall i,j$.

The traction terms in the rotated system can be obtained with
\begin{equation}
	\tens{\Lambda}^T \tens{\sigma} \tens{\Lambda} = 
	\begin{bmatrix}
		\nhat{} \cdot \tens{\sigma} \nhat{} & \nhat{} \cdot \tens{\sigma} \that{} & \nhat{} \cdot \tens{\sigma} \thati{}\\
		\that{} \cdot \tens{\sigma} \nhat{} & \that{} \cdot \tens{\sigma} \that{} & \that{} \cdot \tens{\sigma} \thati{}\\
		\thati{} \cdot \tens{\sigma} \nhat{} & \thati{} \cdot \tens{\sigma} \that{} & \thati{} \cdot \tens{\sigma} \thati{}
	\end{bmatrix}
	= 
	\tens{\Lambda}^T \tens{\tau} \tens{\Lambda}
	- p\tens{I}
	.
\label{eq:sigma-rot}
\end{equation}
Thus the traction constraints to be applied on the boundary segments $\Gamma_S$ are given by
\begin{equation}
	\tens{\mathcal{H}} \odot \left( \tens{\Lambda}^T \tens{\tau}_S \tens{\Lambda} \right) \equiv \tens{G}_S \quad \forall \vec x \in \Gamma_S
\label{eq:stress-data}
\end{equation}
with $\tens{\tau}_S$ an arbitrarily defined deviatoric stress tensor containing the stress constraints.
Note that because the velocity is constrained in the $\nhat{}$ direction, we require that $\mathcal{H}_{00} = 0$.
Moreover, due to the boolean nature of $\tens{\mathcal{H}}$ the stress along boundary can be decomposed as:
\begin{equation}
	\begin{split}
		\tens{\Lambda}^T \tens{\sigma} \tens{\Lambda} &= \left( \tens{1} - \tens{\mathcal{H}} \right) \odot \left( \tens{\Lambda}^T \tens{\sigma} \tens{\Lambda} \right) + \tens{\mathcal{H}} \odot \left( \tens{\Lambda}^T \tens{\sigma} \tens{\Lambda} \right) \\
		&= \left( \tens{1} - \tens{\mathcal{H}} \right) \odot \left( \tens{\Lambda}^T \tens{\tau} \tens{\Lambda} \right) + \tens{G}_S - p \tens{I}
	\end{split}
\label{eq:stress-rot}
\end{equation} 
where $\tens{G}_S$ is the tensor containing the stress constraints to be applied.

Along the boundary $\Gamma_S$, the traction vector can be defined by
\begin{equation}
	\begin{split}
		\tens{\sigma} \nb{} = \tens{\Lambda} \left( \tens{\Lambda}^T \tens{\sigma} \tens{\Lambda} \right) \tens{\Lambda}^T \nb{} = \sum_i \sum_j \left(\nb{} \cdot \vec{\Lambda}_i \right) \left( \vec{\Lambda}_j \cdot \tens{\sigma} \vec{\Lambda}_i \right) \vec{\Lambda}_j . 
	\end{split}
\label{eq:sigma-n}
\end{equation}
Therefore, combining Eqs.~\eqref{eq:stress-rot}~\&~\eqref{eq:sigma-n} we obtain
\begin{equation}
	\begin{split}
		\tens{\sigma} \nb{} &= \tens{\Lambda} \left[ \left( \tens{1} - \tens{\mathcal{H}} \right) \odot \left( \tens{\Lambda}^T \tens{\tau} \tens{\Lambda} \right) + \tens{\mathcal{H}} \odot \left( \tens{\Lambda}^T \tens{\tau} \tens{\Lambda} \right) \right]\tens{\Lambda}^T \nb{} - p \nb{} \\
		&= \sum_i \sum_j \left(\nb{} \cdot \vec{\Lambda}_i \right) \left( \left( 1-\mathcal{H}_{ij} \right) \vec{\Lambda}_j \cdot \tens{\tau} \vec{\Lambda}_i \right) \vec{\Lambda}_j + \left(\nb{} \cdot \vec{\Lambda}_i \right) \tens{G}_{S_{ij}} \vec{\Lambda}_j - p \nb{}
	\end{split}
\end{equation}
on $\Gamma_S$.

\section{Weak form} 
\label{sec:weak_form}
In the domain $\Omega$ we employ a mixed finite element formulation to solve equations~\eqref{eq:momentum}~\&~\eqref{eq:div}. 
We define the velocity space
\begin{equation}
	\vec{H}_0^1(\Omega) := \left \{\vec v \in \vec{H}^1 (\Omega) : \vec{v}|_{\Gamma_D} = \vec 0 \right\},
\end{equation}
and the pressure space
\begin{equation}
	L_0^2(\Omega) := \left \{ q \in L^2(\Omega) : \int_{\Omega} q = 0 \right \}.
\end{equation}
Where $\vec v \in \vec{H}_0^1$ and $q \in L_0^2$ are the test functions associated with the velocity $\vec u$ and the pressure $p$ respectively. 
Multiplying Eq.~\eqref{eq:momentum} by $\vec v$ and Eq.~\eqref{eq:div} by $q$ and integrating by part yields the weak form of the Stokes problem
\begin{equation}
	\mathcal{A} \left( (\vec v, q);(\vec u, p) \right) = \mathcal{F} (\vec v), \quad \forall (\vec v,q) \in \vec{H}_0^1(\Omega) \times L_0^2(\Omega)
\end{equation}
where the bilinear form is given by
\begin{equation}
	\begin{split}
		\mathcal{A} \left( \left(\vec v, q \right); \left(\vec u, p \right) \right) &:= 
		\mathcal{A}_V (\vec v, \vec u, \eta(\vec u, p))_{\Omega} - \mathcal{B} (\vec u, q)_{\Omega} - \mathcal{B} (\vec v, p)_{\Omega}\\
		& - \mathcal{A}_S (\vec v, \vec u, \eta(\vec u, p))_{\Gamma_S} + \mathcal{A}_p (\vec v, p)_{\Gamma_S}
	\end{split}
\end{equation}
with 
\begin{align}
	\mathcal{A}_V (\vec v, \vec u, \eta(\vec u, p) )_{\Omega} &:= 
	\int_{\Omega} 2 \eta (\vec u, p) \tens{\varepsilon} (\vec u) : \tens{\varepsilon} (\vec v) \,dV
	\label{eq:form-stress-volume} \\
	\mathcal{B} (\vec v, p)_{\Omega} &:=
	\int_{\Omega} p \nabla \cdot \vec v \, dV
	\label{eq:form-div-u} \\
	\mathcal{A}_S (\vec v, \vec u, \eta(\vec u, p))_{\Gamma_S} &:= 
	\sum_i \sum_j \int_{\Gamma_S} 
	\left(\vec v \cdot \vec{\Lambda}_i \right)
	\left(\nb{} \cdot \vec{\Lambda}_j \right)
	\left( 1-\mathcal{H}_{ij} \right) \vec{\Lambda}_i \cdot \left(2 \eta(\vec u, p)\tens{\varepsilon}(\vec u)\right) \vec{\Lambda}_j \, dS
	\label{eq:form-1minusH} \\
	\mathcal{A}_p (\vec v, p)_{\Gamma_S} &:= \int_{\Gamma_S} \vec v \cdot p \nb{} \, dS
	\label{eq:form-surf-p}
\end{align}
and the linear form is given by
\begin{equation}
	\mathcal{F} (\vec v) := \mathcal{F}_V (\vec v)_{\Omega}
	+ \mathcal{F}_S (\vec v)_{\Gamma_S}
\end{equation}
where
\begin{align}
	\mathcal{F}_V (\vec v)_{\Omega} &:= 
	\int_{\Omega} \vec v \cdot \rho \vec g \, dV \\
	\mathcal{F}_S (\vec v)_{\Gamma_S} &:= 
	\sum_i \sum_j \int_{\Gamma_S}
	\left(\vec v \cdot \vec{\Lambda}_i \right)
	\left(\vec n \cdot \vec{\Lambda}_j \right)
	\left(G_S \right)_{ij} \, dS
	\label{eq:form-H}
\end{align}

\section{Nitsche's method} 
\label{sec:nitsche_method}

To weakly impose the slip constraint we use Nitsche's method \citep{Nitsche1971}. 
To that end, we first define the discrete spaces $\vec v^h , \vec u^h \subset \vec{H}_0^1$ 
and $p^h , q^h \subset L_0^2$.
Then, we introduce the penalty of the slip constraint.
Its bilinear form is defined as
\begin{equation}
	\mathcal{A}_{\gamma}^h \left(\vec v^h, \vec u^h \right)_{\Gamma_S} := \int_{\Gamma_S} \gamma \left(\vec v^h \cdot \vec{\Lambda}_0 \right) \left(\vec u^h \cdot \vec{\Lambda}_0 \right) \, dS
\label{eq:nitsche-bilinear}
\end{equation}
and its linear form as
\begin{equation}
	\mathcal{F}_{\gamma}^h (\vec v^h)_{\Gamma_S} := \int_{\Gamma_S} \gamma \left(\vec v^h \cdot \vec{\Lambda}_0 \right) \bar{g} \quad dS.
\label{eq:nitsche-linear}
\end{equation}
The bilinear form of the discrete weak form is thus defined as
\begin{equation}
	\mathcal{A}^h \left( \left(\vec v^h, q^h \right); \left(\vec u^h, p^h \right) \right) :=
	\mathcal{A} \left( \left(\vec v^h, q^h \right); \left(\vec u^h, p^h \right) \right)
	+ \mathcal{A}_{\gamma}^h \left(\vec v^h, \vec u^h \right)_{\Gamma_S}
\end{equation}
and the linear form as
\begin{equation}
	\mathcal{F}^h (\vec v^h) := 
	\mathcal{F} (\vec v^h) 
	+ \mathcal{F}_{\gamma}^h (\vec v^h)_{\Gamma_S}.
\end{equation}

The penalty parameter $\gamma$ should be large enough to ensure that the bilinear form is coercive.
Minimal values for $\gamma$ at each facet $f$ were proposed for triangular/tetrahedral meshes \citep{Shahbazi2005}
\begin{equation}
	\gamma_f > \frac{k (k+d-1) A_f}{d V_f},
\end{equation}
and quadrilateral/hexahedral meshes \citep{Hillewaert2013}
\begin{equation}
	\gamma_f > (k+1)^2 \frac{A_f}{V_f},
\end{equation}
with $k$ the polynomial order of the finite element velocity space, $d$ the number of spatial dimensions, $A_f$ the area of the facet and $V_f$ the volume of the cell containing the facet.

In addition, a symmetrising term is added to enhance stability of the discrete weak form.
The symmetry term acts only on the weakly imposed Dirichlet constraint, i.e. on the $\nhat{}$ ($\vec{\Lambda_0}$) component.
Moreover, in the bilinear form defined by Eq~\eqref{eq:form-1minusH} the only component that cannot be constrained by data in the linear form defined by Eq.~\eqref{eq:form-H} is the component involving $\nhat{} \cdot \tens{\tau} \nhat{}$.
Thus the symmetric form only applies to that component such that in the bilinear form it is defined by the terms
\begin{align}
	\mathcal{A}_S^S (\vec v^h, \vec u^h, \eta^h(\vec u^h, p^h))_{\Gamma_S} &:= 
	\int_{\Gamma_S} \left(\vec u^h \cdot \vec{\Lambda}_0 \right) \left(\nb{} \cdot \vec{\Lambda}_0 \right) \vec{\Lambda}_0 \cdot \left(2 \eta^h (\vec u^h, p^h)\tens{\varepsilon}(\vec v^h) \right) \vec{\Lambda}_0 \, dS
	\label{eq:form-symmetry-v} \\
	\mathcal{B}_S^S(\vec u^h, q^h)_{\Gamma_S} &:= 
	- \int_{\Gamma_S} \left(\vec u^h \cdot \vec{\Lambda}_0 \right) \left(\nb{} \cdot \vec{\Lambda}_0 \right) q^h \, dS
	\label{eq:form-symmetry-q}
\end{align}
and in the linear form by
\begin{align}
	\mathcal{F}_{S_{\vec v}}^S (\vec v^h, \eta^h(\vec u^h, p^h))_{\Gamma_S} &:= 
	\int_{\Gamma_S} \bar{g} (\nb{} \cdot \vec{\Lambda}_0) \vec{\Lambda}_0 \cdot \left(2 \eta^h (\vec u^h, p^h)\tens{\varepsilon}(\vec v^h) \right) \vec{\Lambda}_0 \, dS \\
	\mathcal{F}_{S_{q}}^S (q^h)_{\Gamma_S} &:= 
	- \int_{\Gamma_S} \bar{g}\left(\nb{} \cdot \vec{\Lambda}_0 \right) q^h \, dS.
\end{align}

Finally, the complete discrete weak form for generalised Navier-slip conditions is given by
\begin{equation}
	\begin{split}
		&\mathcal{A}_V (\vec v^h, \vec u^h, \eta(\vec u^h, p^h))_{\Omega} 
		- \mathcal{B} (\vec u^h, q^h)_{\Omega} 
		- \mathcal{B} (\vec v^h, p^h)_{\Omega} 
		- \mathcal{A}_S (\vec v^h, \vec u^h, \eta(\vec u^h, p^h))_{\Gamma_S} 
		+ \mathcal{A}_p (\vec v^h, p^h)_{\Gamma_S} \\
		& - \mathcal{A}_S^S (\vec v^h, \vec u^h, \eta(\vec u^h, p^h))_{\Gamma_S} 
		- \mathcal{B}_S^S(\vec u^h, q^h)_{\Gamma_S} 
		+ \mathcal{A}_{\gamma}^h \left(\vec v^h, \vec u^h \right)_{\Gamma_S} \\
		&= \mathcal{F} (\vec v^h)_{\Omega} 
		+ \mathcal{F}_S (\vec v^h)_{\Gamma_S} 
		- \mathcal{F}_{S_{\vec v}}^S (\vec v^h, \eta^h(\vec u^h, p^h))_{\Gamma_S} 
		- \mathcal{F}_{S_{q}}^S (q^h)_{\Gamma_S} 
		+ \mathcal{F}_{\gamma}^h (\vec v^h)_{\Gamma_S}
	\end{split}
\label{eq:complete-weak-form}
\end{equation}

\section{Residual form} 
\label{sec:residual_form}
To solve the problem described by Eq.~\eqref{eq:complete-weak-form} with iterative methods we express the problem in term of residual.
First, using the finite element method to discretize Eq.~\eqref{eq:complete-weak-form} we define
\begin{equation*}
	\begin{matrix}
		\vec u^h = \sum_i \boldsymbol{\phi}_i \vec{u}_i, & &
		p^h = \sum_i \varphi_i p_i, & &
		\vec v^h = \sum_i \boldsymbol{\psi}_i \vec{v}_i, & &
		q^h = \sum_i \zeta_i q_i
	\end{matrix}
\end{equation*}
with $\vec u^h$, $p^h$ the discrete solution for velocity and pressure respectively, $\boldsymbol{\phi}_i$, $\boldsymbol{\psi}_i$ the element basis functions and $\vec{u}_i$, $p_i$ the coefficient of the trial functions for velocity and pressure respectively. The same decomposition holds for the test function with $\vec v^h$, $q^h$ the discrete test functions for velocity and pressure respectively, $\boldsymbol{\psi}_i$, $\zeta_i$ the element basis functions and $\vec{v}_i$, $q_i$ the coefficient of the test functions for velocity and pressure respectively.

\subsection{Linear residual} 
\label{sub:linear_residual}
From the weak form defined at Eq~\eqref{eq:complete-weak-form} the linear residual for velocity ($\vec{R_u}$) and pressure ($\vec{R_p}$) can be written as
\begin{equation}
	\begin{bmatrix}
		\vec{R_u}\\
		\vec{R_p}
	\end{bmatrix}
	=
	\begin{bmatrix}
		\tens{A}(\boldsymbol{\psi},\boldsymbol{\phi}) \vec u 
		+ \tens{B}(\boldsymbol{\psi},\varphi) \vec p 
		- \vec{f_u}(\boldsymbol{\psi})\\
		\tens{C}(\boldsymbol{\phi},\zeta)\vec u 
		- \vec{f_p}(\zeta)
	\end{bmatrix}
\label{eq:linear-residual}
\end{equation}
where
\begin{align}
	A_{ij}(\boldsymbol{\psi}_i,\boldsymbol{\phi}_j) &:= 
	\mathcal{A}_V (\boldsymbol{\psi}_i,\boldsymbol{\phi}_j)_{\Omega} 
	- \mathcal{A}_S (\boldsymbol{\psi}_i,\boldsymbol{\phi}_j)_{\Gamma_S} 
	- \mathcal{A}_S^S (\boldsymbol{\psi}_i, \boldsymbol{\phi}_j)_{\Gamma_S} 
	+ \mathcal{A}_{\gamma}^h \left(\boldsymbol{\psi}_i,\boldsymbol{\phi}_j \right)_{\Gamma_S},\\
	B_{ij} (\boldsymbol{\psi}_i,\varphi_j) &:= 
	- \mathcal{B} (\boldsymbol{\psi}_i,\varphi_j)_{\Omega}
	+ \mathcal{A}_p (\boldsymbol{\psi}_i,\varphi_j)_{\Gamma_S},\\
	C_{ij} (\boldsymbol{\phi}_i,\zeta_j) &:= 
	- \mathcal{B} (\boldsymbol{\phi}_i,\zeta_j)_{\Omega}
	- \mathcal{B}_S^S(\boldsymbol{\phi}_i,\zeta_j)_{\Gamma_S}
\end{align}
and
\begin{align}
	f_{\vec{u}_i} (\boldsymbol{\psi}_i) &:= 
	\mathcal{F} (\boldsymbol{\psi}_i)_{\Omega} 
	+ \mathcal{F}_S (\boldsymbol{\psi}_i)_{\Gamma_S} 
	- \mathcal{F}_{S_{\vec v}}^S (\boldsymbol{\psi}_i)_{\Gamma_S}
	+ \mathcal{F}_{\gamma}^h (\boldsymbol{\psi}_i)_{\Gamma_S}\\
	f_{p_i}(\zeta_i) &:= - \mathcal{F}_{S_{q}}^S (\zeta_i)_{\Gamma_S}.
\end{align}

Note that for the volumetric part $\tens{C} = \tens{B}^T$ and that the symmetry of the saddle point problem is conserved.

\subsection{Non-linear residual} 
\label{sub:non_linear_residual}
As introduced in Eq.~\eqref{eq:tau}, common Earth Sciences problems involve non-linear Arrhenius viscosity laws that can depend on both the velocity ($\vec u$) and pressure ($p$) e.g., Eq.~\eqref{eq:Arrh-viscosity}.
To solve these non-linearities using iterative method, at iteration $k+1$ with the previous guess $(\vec u_k, p_k)$ we define the non-linear residual
\begin{equation}
	\begin{bmatrix}
		\vec{R}_{\vec u}^k\\
		\vec{R}_{\vec p}^k
	\end{bmatrix}
	=
	\begin{bmatrix}
		\tens{A}(\boldsymbol{\psi},\boldsymbol{\phi},\eta(\vec u^k,p^k)) \vec u^k 
		+ \tens{B}(\boldsymbol{\psi},\varphi) \vec p^k 
		- \vec{f_u}(\boldsymbol{\psi})\\
		\tens{C}(\boldsymbol{\phi},\zeta)\vec u^k 
		- \vec{f_p}(\zeta)
	\end{bmatrix}
\label{eq:non-linear-residual}
\end{equation}
with
\begin{equation}
	\begin{split}
		A_{ij} (\boldsymbol{\psi}_i,\boldsymbol{\phi}_j,\eta(\vec u^k,p^k)) &:=
		\mathcal{A}_V (\boldsymbol{\psi}_i,\boldsymbol{\phi}_j,\eta(\vec u^k,p^k))_{\Omega} 
		- \mathcal{A}_S (\boldsymbol{\psi}_i,\boldsymbol{\phi}_j,\eta(\vec u^k,p^k))_{\Gamma_S} \\
		&- \mathcal{A}_S^S (\boldsymbol{\psi}_i, \boldsymbol{\phi}_j,\eta(\vec u^k,p^k))_{\Gamma_S} 
		+ \mathcal{A}_{\gamma}^h \left(\boldsymbol{\psi}_i,\boldsymbol{\phi}_j \right)_{\Gamma_S}
	\end{split}
\end{equation}

\section{Benchmarking} 
\label{sec:benchmarking}
\subsection{2D Rotated Couette Flow} 
\label{sub:2d_rotated_couette_flow}

\begin{figure}[h!]
    \centering
    \includegraphics[scale=0.9]{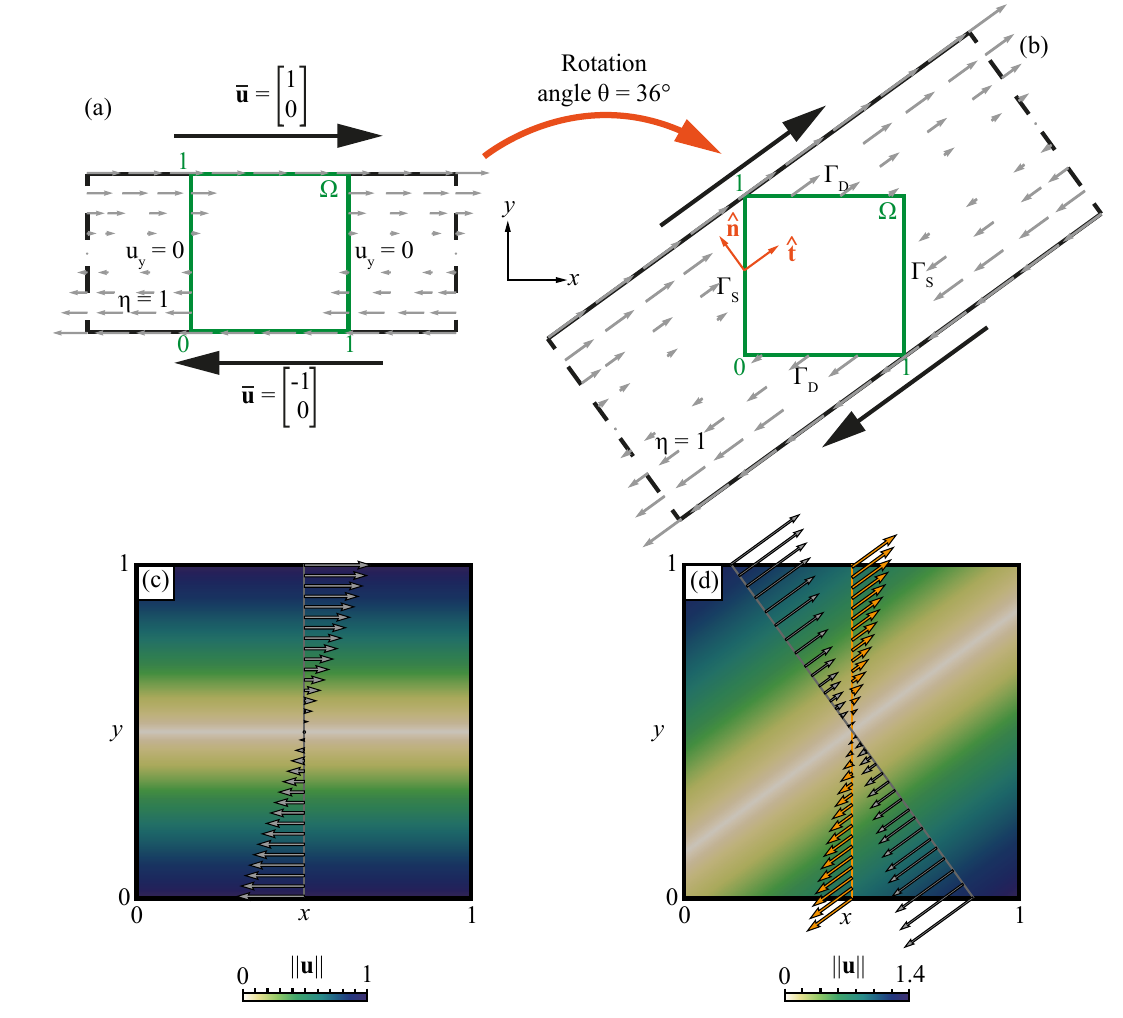}
    \caption{(a) Couette flow setup. 
    The black lines represent two plates moving relatively to each others.
    The green square represent the domain $\Omega$ in which we solve for $\vec u$.
    The grey arrows show the flow field outside the domain $\Omega$.
    (b) Rotation of the system represented in (a) of an angle $\theta = 36^{\circ}$.
    The grey arrows show the rotated flow field outside the domain $\Omega$.
    The orange arrows show a representation of $\vec{\hat{t}}$ and $\nhat{}$, the special direction describing the Navier-slip boundary conditions.
    (c) Solution for $\vec u$ in the non-rotated system.
    The grey arrows show the velocity field along the vertical section.
    (d) Solution for $\vec u$ in the rotated system.
    The grey arrows show the velocity field along a section oriented at $\theta = 36^{\circ}$.
    The orange arrows show the velocity field along the vertical section located at $x=0.5$.}
    \label{fig:couette-flow}
\end{figure}

Couette flow describes the flow between two plates moving relatively to each others (Figure~\ref{fig:couette-flow}a~\ref{fig:couette-flow}c).
In the following experiments we use a constant viscosity $\eta = 1$.
First, we consider a domain $\Omega = [0,1]^2$ with the following boundary conditions:
\begin{align}
	\vec u &= 
	\begin{bmatrix}
		-1\\
		0 	
	\end{bmatrix} 
	&\forall y = 0\\
	\vec u &=  
	\begin{bmatrix}
		1\\
		0 	
	\end{bmatrix} 
	&\forall y = 1\\
	u_y &= 0 &\forall x = \{0,1\}
\end{align}
Without pressure gradient imposed, such conditions reduce the conservation of momentum to a single variable problem:
\begin{equation}
	\frac{\partial^2 \vec u}{\partial y^2} = \vec 0.
\label{eq:couette-2d}
\end{equation}
Integrating twice and using the boundary conditions to determine the integration constants gives
\begin{equation}
	\vec u =  
	\begin{bmatrix}
		2 y - 1\\
		0
	\end{bmatrix}
	.
\label{eq:couette-solution}
\end{equation}
Rotating the whole system by an angle $\theta$ (Figure~\ref{fig:couette-flow}b) while staying in the same coordinate system transforms $\vec u$ to
\begin{equation}
	\vec{u}_R(\vec x) = \tens{R} (\theta) \vec u \left(\tens{R}^T (\theta) \vec x \right) 
\label{eq:rotated-couette}
\end{equation}
with the rotation matrix given by
\begin{equation}
	\tens{R} (\theta) = 
	\begin{bmatrix}
		\cos \theta & -\sin \theta\\
		\sin \theta & \cos \theta 		
	\end{bmatrix} 	
	.
\end{equation} 

Numerically solving the rotated Couette flow with generalised Navier-slip boundary conditions using Nitsche's method must verify the same solution than the solution given by Eq.~\eqref{eq:rotated-couette}.
Thus, we consider a domain $\Omega = [0,1]^2$ crossed by the rotated Couette flow (Figure~\ref{fig:couette-flow}b).
To numerically solve this configuration we impose the following boundary conditions:
\begin{align}
	\vec u &= \vec{u}_R(\vec x) & \forall \vec x \in \Gamma_D\\
	\vec u \cdot \nhat{} &= 0 & \forall \vec x \in \Gamma_S\\
	\tens{G}_S &= \tens{\mathcal{H}} \odot \left(\tens{\Lambda}^T \tens{\tau}_R \tens{\Lambda} \right) & \forall \vec x \in \Gamma_S
\end{align}
with 
\begin{equation}
	\tens{\mathcal{H}} = 
	\begin{bmatrix}
		0 & 1\\
		1 & 0
	\end{bmatrix}
	,
\end{equation}
\begin{equation}
	\tens{\Lambda} = 
	\begin{bmatrix}
		\vec{\Lambda_0} & \vec{\Lambda_1}
	\end{bmatrix}
	=
	\begin{bmatrix}
		\nhat{} & \vec{\hat{t}}
	\end{bmatrix}
\end{equation}
and
\begin{align}
	\vec{\hat{t}} &= \tens{R} (\theta) 
	\begin{bmatrix}
		1\\
		0
	\end{bmatrix}\\
	\nhat{} &= \tens{R} \left(\frac{\pi}{2} \right) \vec{\hat{t}}.
\end{align}
To obtain the stress $\tens{\tau}_R$ we first rotate the constant strain rate tensor $\tens{\varepsilon}$:
\begin{equation}
	\tens{\varepsilon}_R = \tens{R} (\theta) \, \tens{\varepsilon}\, \tens{R}^T (\theta)
\end{equation}
and compute
\begin{equation}
	\tens{\tau}_R = 2 \eta \tens{\varepsilon}_R.
\end{equation}

Figure~\ref{fig:couette-flow}d shows the solution of the rotated Couette flow using Nitsche's method to impose Navier-slip boundary conditions along $\Gamma_S$.
The solution obtained exactly reproduces the analytical solution of the rotated flow down to machine precision error.

\subsection{3D oblique stretching} 
\label{sub:3d_oblique_stretching}

The following experiments aim to demonstrate the application of the generalised Navier-slip boundary conditions to the scale of 3D geodynamics models.

\begin{figure}[h!]
    \centering
    \includegraphics[scale=0.9]{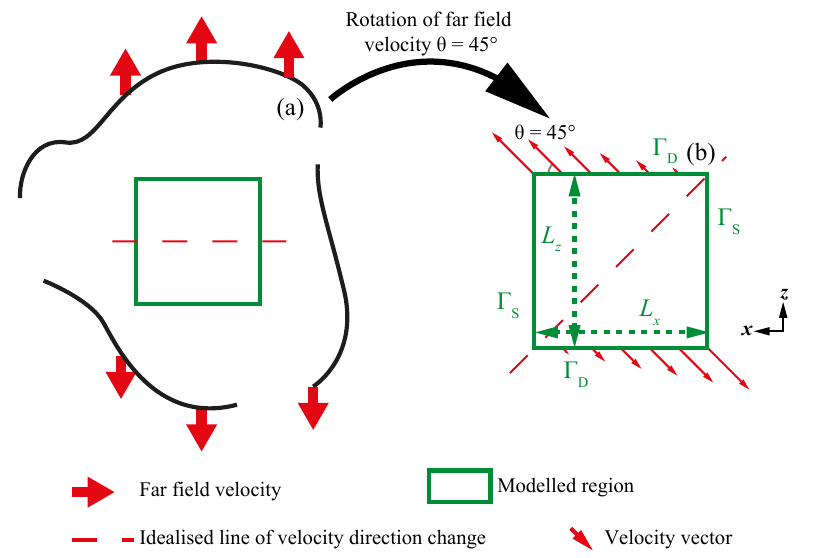}
    \caption{Map view representation of a far field velocity field containing the region of interest to be modelled.
    (a) Pure extension in $\boldsymbol z$ direction, constant velocity along $\boldsymbol x$ direction.
    (b) Rotated velocity field with an angle $\theta=45^{\circ}$, velocity vectors are rotated but not the regional domain.
    $\Gamma_D$ and $\Gamma_S$ represent the Dirichlet and the Navier-slip boundaries respectively.
    $L_x$ and $L_z$ represent the length of the domain in $\boldsymbol x$ and $\boldsymbol z$ directions respectively.
    }
    \label{fig:setup-pure-extension}
\end{figure}

We consider a domain $\Omega = [0,1000]^3$ km$^3$ with a constant density $\rho = 3300$ kg.m$^3$ and a gravity acceleration vector $\vec g = [0, -9.8, 0]$ m.s$^{-2}$.
We provide experiments using two sets of boundary conditions (Figure~\ref{fig:setup-pure-extension} and Figure~\ref{fig:setup-oblique-extension}).
Each set of boundary conditions is applied to a constant viscosity domain and a variable viscosity domain especially designed as an extreme case implying stress discontinuity between faces.

\subsubsection{Rotated pure extension} 
\label{ssub:rotated_pure_extension}

First we consider a simple case of pure extension in the $x-z$ plane (Figure~\ref{fig:setup-pure-extension}a) with a vertical velocity inflow at the base and a free surface:
\begin{align}
	\bar{u}_z &= 1  & \forall x = L_x             \label{eq:extension-xmax}\\
	\bar{u}_z &= -1 & \forall x = 0               \label{eq:extension-xmin}\\
	\bar{u}_x &= 0 & \forall z = \{0,L_z\}        \label{eq:extension-z}\\
	\bar{u}_y &= 1 & \forall y = 0                \label{eq:extension-ymin}\\
	\tens{\sigma}\nb{} &= \vec 0 &\forall y = L_y \label{eq:extension-ymax}
\end{align}
To obtain an oblique extension case we rotate this system of an angle $\theta$ in the horizontal ($x-z$) plane (Figure~\ref{fig:setup-pure-extension}) using the rotation matrix
\begin{equation}
	\tens{R}_{y}(\theta) =
	\begin{bmatrix}
		\cos \theta & 0 & \sin \theta\\
		0 & 1 & 0\\
		-\sin \theta & 0 & \cos \theta
	\end{bmatrix}
\end{equation}
while keeping the same coordinate system and the same domain coordinates i.e., $\Omega$ is not rotated (Figure~\ref{fig:setup-pure-extension}).
In this new configuration we require to impose Dirichlet boundary conditions on faces of normal $\boldsymbol z$ denoted by $\Gamma_D$ and Navier-slip boundary conditions on faces of normal $x$ denoted by $\Gamma_S$.
The top and bottom boundary conditions are not modified.
To obtain Dirichlet boundary conditions we reduce the problem to a 2D case of pure extension that we rotate.
This choice implies that the vertical motions do not influence the horizontal components of the flow field.
Thus, considering only the horizontal components of the velocity with the boundary conditions \eqref{eq:extension-xmax}-\eqref{eq:extension-z} the non rotated flow field satisfies
\begin{equation}
	\frac{\partial^2 \vec u}{\partial z^2} = \vec 0.
\end{equation}
Therefore, integrating twice and using the boundary conditions to determine the constants of integration yields
\begin{equation}
	\vec u = \frac{2}{L_z}
	\begin{bmatrix}
		0\\
		0\\
		z - 1
	\end{bmatrix}	
	.
\end{equation}
The rotation of the referential requires to compute a new velocity field $\vec{u}_R$ in that referential with Eq.~\eqref{eq:rotated-couette} using $\tens{R}_{y}(\theta)$.
Thus, to numerically solve the 3D oblique stretching model we apply the following boundary conditions:
\begin{align}
	\bar{u}_x &= u_{R_x} &\forall z = \{0,L_z\}\\
	\bar{u}_z &= u_{R_z} &\forall z = \{0,L_z\}\\
	\vec u \cdot \nhat{} &= 0 &\forall x = \{0,L_x\}\\
	\tens{G}_S &= \tens{\mathcal{H}} \odot \left(\tens{\Lambda}^T \tens{\tau}_R \tens{\Lambda}\right) &\forall x = \{0,L_x\}
\end{align}
with
\begin{equation}
	\begin{matrix}
		\that{} = \tens{R}_{y}(\theta)
		\begin{bmatrix}
			0\\
			0\\
			1
		\end{bmatrix}
		, \quad
		& \nhat{} = \tens{R}_{y}(-\frac{\pi}{2}) \that{}, \quad
		& \thati{} = \nhat{} \times \that{},
	\end{matrix}	
\end{equation}
\begin{equation}
	\tens{\Lambda} = 
	\begin{bmatrix}
		\vec{\Lambda}_0 & \vec{\Lambda}_1 & \vec{\Lambda}_2
	\end{bmatrix}
	=
	\begin{bmatrix}
		\nhat{} & \that{} & \thati{}
	\end{bmatrix}
\label{eq:Lambda-pure-extension}
\end{equation}
and
\begin{equation}
	\tens{\mathcal{H}} = 
	\begin{bmatrix}
		0 & 1 & 1\\
		1 & 1 & 0\\
		1 & 0 & 0
	\end{bmatrix}
\label{eq:H-pure-extension}
\end{equation}

To compute $\tens{\tau}_R$ we define a background virtual flow field based on the Dirichlet boundary forcing. 
We then use this virtual flow field to compute a background strain rate $\tens{\varepsilon}_b (\vec u)$
that we use to compute the stress boundary values with
\begin{equation}
	\tens{\tau}_R = 2 \eta \left(\tens{R}_y (\theta) \tens{\varepsilon}_b \tens{R}_y^T (\theta) \right)
\end{equation}

\begin{figure}[h!]
    \centering
    \includegraphics[scale=0.9]{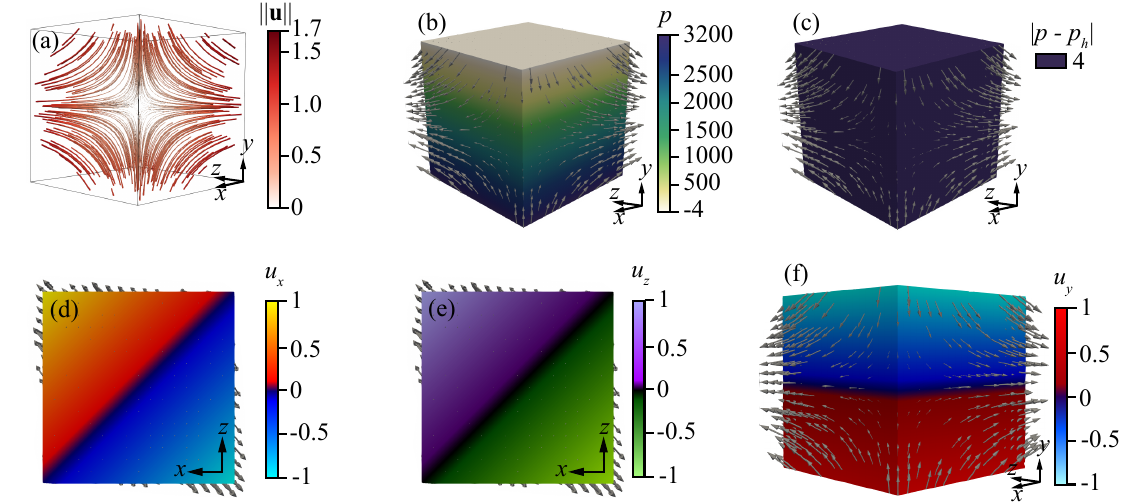}
    \caption{
    (a) Streamlines of the flow field solution, the colour shows the value of $\| \vec u\|$ in m.s$^{-1}$.
    (b) Total pressure $p$ solution $\times 10^7$ Pa.
    (c) Dynamic pressure $|p-p_h| \times 10^7$ Pa with $p_h$ the hydrostatic pressure.
    Values of the components of the velocity field: (d) $u_x$, (e) $u_z$, (f) $u_y$ in m.s$^{-1}$.}
    \label{fig:const-eta-45}
\end{figure}

Figure~\ref{fig:const-eta-45} shows the result of this model with a constant viscosity $\eta = 10^{21}$ Pa.s.
With the rotation angle $\theta = 45^{\circ}$ the horizontal flow field is symmetric with respect to the diagonal of the domain (Figure~\ref{fig:const-eta-45}a, \ref{fig:const-eta-45}d, \ref{fig:const-eta-45}e). 
The total pressure field (Figure~\ref{fig:const-eta-45}b) shows an increase in depth while the dynamic pressure is constant over the domain with a value of $2.8 \times 10^{6}$ Pa (Figure~\ref{fig:const-eta-45}c).
Using the same parameters but changing the rotation angle $\theta$ does not affect the pressure field.
The vertical component of the flow field (Figure~\ref{fig:const-eta-45}f) is also kept unchanged with respect to the rotation.

\begin{figure}[h!]
    \centering
    \includegraphics[scale=0.9]{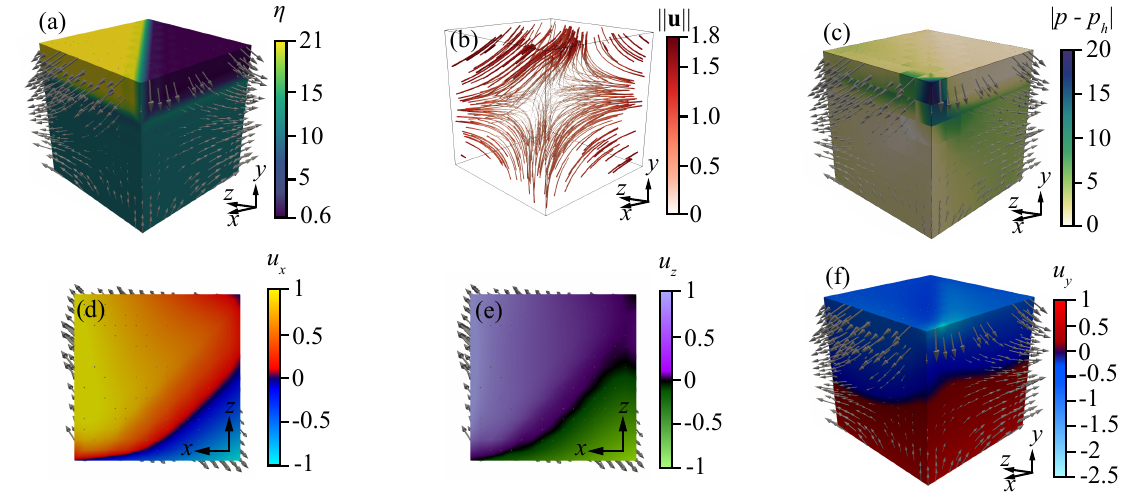}
    \caption{
    (a) Viscosity field $\eta \times 10^{22}$ Pa.s.
    (b) Streamlines of the flow field solution, the colour shows the value of $\| \vec u\|$ in m.s$^{-1}$.
    (c) Dynamic pressure $|p-p_h| \times 10^7$ Pa with $p_h$ the hydrostatic pressure.
    Values of the components of the velocity field: (d) $u_x$, (e) $u_z$, (f) $u_y$ in m.s$^{-1}$.}
    \label{fig:variable-eta-45}
\end{figure}

Figure~\ref{fig:variable-eta-45} shows the result of the $45^{\circ}$ rotated extension with a variable viscosity structure. 
The viscosity structure is imposed with a hyperbolic function from $y=800$ km to $y=1000$ km.
The low and high viscosity regions meet along a vertical diagonal of the domain perpendicularly to the stretching direction (Figure~\ref{fig:variable-eta-45}a) and join the corners of the domain where faces on which we apply Dirichlet and Navier-slip boundary conditions are in contact.
This experiment is designed to obtain a viscosity jump that will enforce a stress discontinuity between the faces of the domain.
Compared to the constant viscosity model the flow field is asymmetric (Figure~\ref{fig:variable-eta-45}b).
The transition zone where velocity directions change occurs inside the low viscosity zone instead of being located in the centre of the domain (Figure~\ref{fig:variable-eta-45}d~\&~\ref{fig:variable-eta-45}e).
As expected the dynamic pressure field shows variations at interfaces between different viscosities regions (Figure~\ref{fig:variable-eta-45}c). 
However, in one of the corners where the low and high viscosity regions meet the stress discontinuity leads to a higher but acceptable pressure value.


\subsubsection{Oblique extension} 
\label{ssub:oblique_extension}

\begin{figure}[h!]
    \centering
    \includegraphics[scale=0.9]{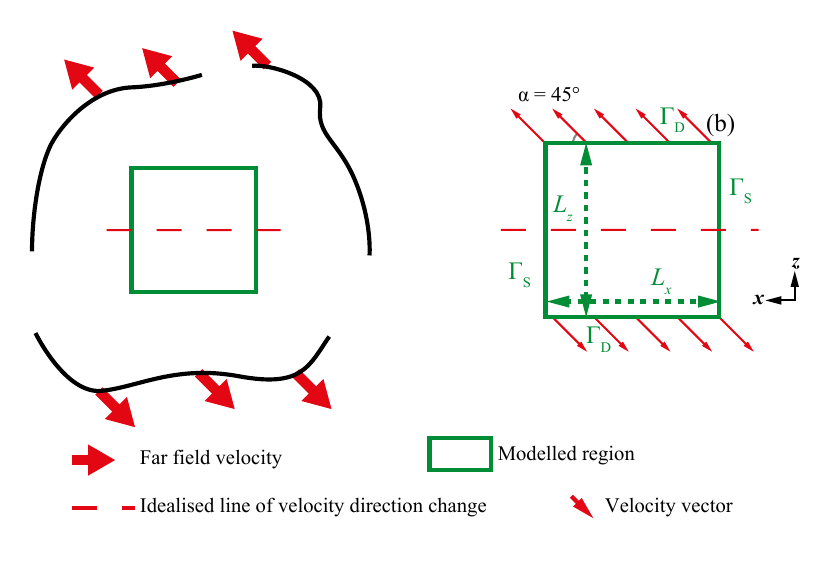}
    \caption{Map view representation of a far field velocity field containing the region of interest to be modelled.
    The velocity field forms an angle of $\alpha = 45^{\circ}$ with the $\boldsymbol x$ and $\boldsymbol z$ directions and has a constant norm along the $\boldsymbol x$ direction.
    $\Gamma_D$ and $\Gamma_S$ represent the Dirichlet and the Navier-slip boundaries respectively.
    $L_x$ and $L_z$ represent the length of the domain in $\boldsymbol x$ and $\boldsymbol z$ directions respectively.
    }
    \label{fig:setup-oblique-extension}
\end{figure}

The second set of experiments models an oblique extension in the $x-z$ plane of an angle $\alpha = 45^{\circ}$ (Figure~\ref{fig:setup-oblique-extension}).
To impose oblique extension we constraint the two horizontal components of the velocity on the Dirichlet boundaries as
\begin{equation}
	\begin{split}
		\bar{u}_{x_0} &= \sqrt{\| \vec{\bar{u}} \|^2 - \bar{u}_{z_0}^2}\\
		\bar{u}_{z_0} &= \| \vec{\bar{u}} \| \cos \alpha
	\end{split}
\end{equation}
with $\| \vec{\bar{u}} \| = 1$ m.s$^{-1}$.
Then we use the following analytical function to compute the boundary velocity value in the required direction:
\begin{equation}
	\vec{\bar{u}} (\vec x) =
	\begin{bmatrix}
		z \frac{2}{L_z} \bar{u}_{x_0} - \bar{u}_{x_0}\\
		0\\
		z \frac{2}{L_z} \bar{u}_{z_0} - \bar{u}_{z_0}
	\end{bmatrix}
\label{eq:analytical-u-oblique}
\end{equation}
The boundary conditions of this model set is thus
\begin{align}
	u_x &= \bar{u}_x & \forall z = \{0, L_z\}\\
	u_z &= \bar{u}_z & \forall z = \{0, L_z\}\\
	\tens{\sigma}\nb{} &= \vec 0 & \forall y = 0\\
	u_y &= \| \vec{\bar{u}} \| & \forall y = L_y\\
	\vec u \cdot \nhat{} &= 0 &\forall x = \{0,L_x\}\\
	\tens{G}_S &= \tens{\mathcal{H}} \odot \left(\tens{\Lambda}^T \tens{\tau}_S \tens{\Lambda} \right) &\forall x = \{0,L_x\}
\end{align}
In the following experiments, we use
\begin{equation}
	\begin{matrix}
		\that{} = 
		\begin{bmatrix}
			\bar{u}_{x_0}\\
			0\\
			\bar{u}_{z_0}
		\end{bmatrix}
		,\quad & \nhat{} = \tens{R}_y(-\frac{\pi}{2}) \that{}
		,\quad & \thati{} = \nhat{} \times \that{},
	\end{matrix}
\end{equation}
while $\tens{\Lambda}$ and $\tens{\mathcal{H}}$ are similar to Eqs.~\eqref{eq:Lambda-pure-extension}~\&~\eqref{eq:H-pure-extension} respectively.

To obtain the imposed stress value $\tens{\tau}_S$ we consider a virtual background strain rate tensor $\tens{\varepsilon}_b (\vec{\bar{u}})$ computed from the analytical function \eqref{eq:analytical-u-oblique} yielding
\begin{equation}
	\tens{\varepsilon}_b (\vec{\bar{u}}) = 
	\frac{1}{L_z}
	\begin{bmatrix}
		0 & 0 & \bar{u}_x\\
    0 & 0 & 0\\
    \bar{u}_x & 0 & 2 \bar{u}_z
	\end{bmatrix}
	.
\label{eq:analytical-eps-oblique}
\end{equation}

\begin{figure}[h!]
    \centering
    \includegraphics[scale=0.9]{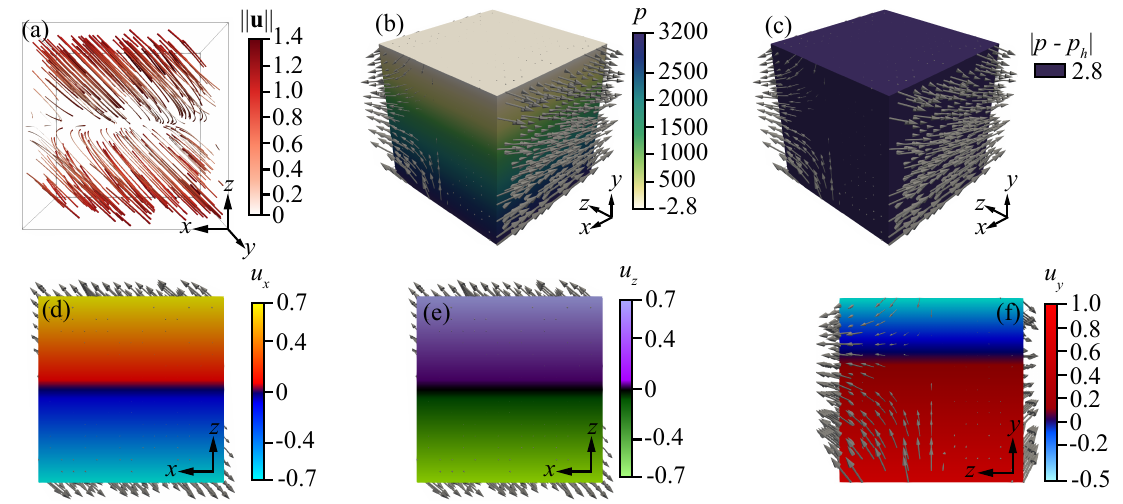}
    \caption{
    (a) Streamlines of the flow field solution, the colour shows the value of $\| \vec u\|$ in m.s$^{-1}$.
    (b) Total pressure $p$ solution $\times 10^7$ Pa.
    (c) Dynamic pressure $|p-p_h| \times 10^7$ Pa with $p_h$ the hydrostatic pressure.
    Values of the components of the velocity field: (d) $u_x$, (e) $u_z$, (f) $u_y$ in m.s$^{-1}$.}
    \label{fig:oblique-constant-eta-45}
\end{figure}

Figure~\ref{fig:oblique-constant-eta-45} shows the model results for a constant viscosity $\eta = 10^{21}$ Pa.s.
The total pressure $p$ (Figure~\ref{fig:oblique-constant-eta-45}b) shows an increase with depth due to the buoyancy forces while the dynamic pressure (Figure~\ref{fig:oblique-constant-eta-45}c) computed as $|p - p_h|$ where $p_h$ is the hydrostatic pressure is constant over the domain with a value of $2.8 \times 10^{6}$ Pa.
The flow field (Figure~\ref{fig:oblique-constant-eta-45}a, \ref{fig:oblique-constant-eta-45}d, \ref{fig:oblique-constant-eta-45}e) shows a localised transition of the horizontal velocity direction in the central part of the domain.
The vertical velocity (Figure~\ref{fig:oblique-constant-eta-45}f) shows an upward flow from the bottom where the Dirichlet boundary condition is applied while the free surface goes downward due to the outflow not being entirely compensated by the bottom inflow.

\begin{figure}[h!]
    \centering
    \includegraphics[scale=0.9]{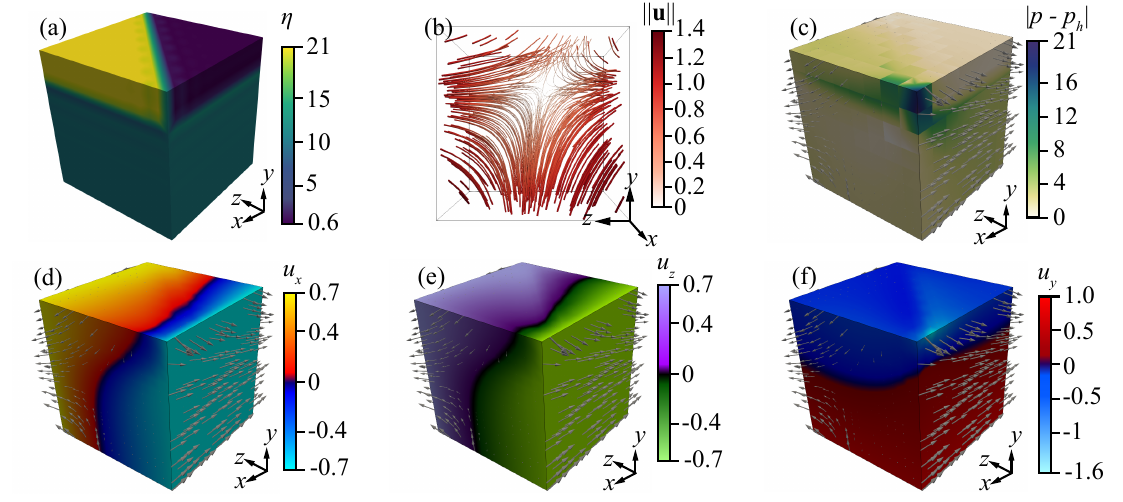}
    \caption{
    (a) Viscosity field $\eta \times 10^{22}$ Pa.s.
    (b) Streamlines of the flow field solution, the colour shows the value of $\| \vec u\|$ in m.s$^{-1}$.
    (c) Dynamic pressure $|p-p_h| \times 10^7$ Pa with $p_h$ the hydrostatic pressure.
    Values of the components of the velocity field: (d) $u_x$, (e) $u_z$, (f) $u_y$ in m.s$^{-1}$.}
    \label{fig:oblique-variable-eta-45}
\end{figure}

Figure~\ref{fig:oblique-variable-eta-45} shows the result of the oblique extension model with a variable viscosity.
We used the same viscosity structure (Figure~\ref{fig:oblique-variable-eta-45}a) than in the rotated extension experiment (Figure~\ref{fig:variable-eta-45}a).
Results are very similar to the rotated extension model (section~\ref{ssub:rotated_pure_extension}) with a variable viscosity (Figure~\ref{fig:variable-eta-45}).
Indeed the flow field is asymmetric and the transition zone where the velocity direction changes is also shifted in the lower viscosity zone (Figure~\ref{fig:oblique-variable-eta-45}b, \ref{fig:oblique-variable-eta-45}d, \ref{fig:oblique-variable-eta-45}e, \ref{fig:oblique-variable-eta-45}f). 
However, the dynamic pressure in the corner is slightly lower than in the previous experiment (Figure~\ref{fig:oblique-variable-eta-45}c).



\section{Geodynamic example: oblique continental rifting}
\label{sec:models-rift}

\subsection{Geodynamic interest}
\label{sub:geodynamic_interest}

Continental rifting is a major ubiquitous process on Earth. 
This process is responsible for the break-up of continents leading to the formation of new oceans.
During continental rifting, the continental crust thins along large scale shear zones, leading to lithospheric and asthenospheric mantle exhumation that will generate the oceanic sea floor, generally by melting and producing oceanic crust.

As shown by several studies \citep[e.g.][]{Brune2018,Jourdon2020a,Jourdon2021}, the tectonic plate motion generally leads toward oblique velocity vectors with respect to the rift system. 
Such obliquity generates non-cylindrical stress field leading to the formation of shear zones accommodating that stress field.
Such a process is known as strain partitioning.
It combines dip-, oblique- and strike-slip shear zones to accomodate the 3D stress field.

A first order question in long term geodynamics is how plate boundaries form and evolve in three dimensions.
Thus, to be able to study this question we need three dimensional approaches.

To show the efficiency of the method we propose and its differences with a more classical approach, we provide two models simulating an oblique extension of 45$^{\circ}$. The first model uses only Dirichlet boundary conditions (Model D) while the second model uses a combination of Dirichlet and generalised Navier-slip boundary conditions to impose the obliquity (Model GNS).

\subsection{Physical model} 
\label{sub:physical_model}

To simulate the long term evolution of the deformation of the lithosphere we use pTatin3D \citep{May2014,May2015} a parallel finite element code that solves the conservation of momentum (Eq.~\ref{eq:momentum}) and mass (Eq.~\ref{eq:div}) for an incompressible fluid. 
The Stokes problem is discretized using $Q_2-P_{1}^{disc}$ basis functions for each element for velocity and pressure respectively.
To model large deformations, an arbitrary Lagrangian-Eulerian approach coupled with the marker-in-cell method \citep{Harlow1965,Sulsky1994} is adopted.

In addition, to take into account temperature variations in time and space during geodynamics processes we solve the following time dependant energy conservation:
\begin{equation}
	\rho_0 C_p \left( \frac{\partial T}{\partial t} + \vec u \cdot \nabla T \right) = \nabla \cdot (k \nabla T) + H
\label{eq:energy}
\end{equation}
with $\rho_0$ the reference density of the material, $C_p$ the thermal heat capacity, $T$ the temperature, $t$ the time, $\vec u$ the velocity of the fluid, $k$ the thermal conductivity and $H$ any heat sources.

Moreover, we use the Boussinesq approximation to vary the density with respect to pressure and temperature as
\begin{equation}
	\rho(p,T) = \rho_0 \left(1 - \alpha(T - T_0) + \beta(p - p_0) \right)
\end{equation}
with $T$ the temperature and $p$ the pressure of the material, $T_0$ and $p_0$ the reference temperature and pressure respectively at which $\rho = \rho_0$ and $\alpha$ and $\beta$ the thermal expansion and compressibility coefficients respectively.

\subsection{Rheological model} 
\label{sub:rheological_model}

The long term rheology of the lithosphere can be assimilated to a high viscosity fluid \citep[e.g.][]{Ranalli1997}.
At low temperatures, the mechanical behaviour of rocks is brittle.
In the following models we use the von Mises plastic yield criterion adapted to continuum mechanics
\begin{equation}
	\eta_p = \frac{C}{2 \varepsilon^{II}},
\label{eq:DP-viscosity}
\end{equation}
to simulate brittle behaviour.
With $C$ the yield stress of the material and
\begin{equation}
  \varepsilon^{II} = \sqrt{ \frac{1}{2} \varepsilon_{ij} \varepsilon_{ij} }
 \label{eq:eps-II}
\end{equation} 
is the square root of the $J_2$ invariant of the strain rate tensor.
In addition, to favour strain localisation and simulate weakening of the rocks in faults we introduce linear plastic strain softening:
\begin{equation}
	C = C_0 - \frac{\epsilon_p - \epsilon_i}{\epsilon_e - \epsilon_i} (C_0 - C_{\infty})
\label{eq:softening}
\end{equation}
with $C_0$ the yield stress of the undamaged material, $\epsilon_p$ the plastic strain, $\epsilon_i = 0$ the amount of plastic strain at which the softening initiates, $\epsilon_e = 1$ the amount of plastic strain at which the softening ends and $C_{\infty}$ the minimum yield stress reached once softening is complete (see parameters in Table~\ref{tab:param}).

At higher temperatures, the rheology of rocks is simulated using non-linear Arrhenius flow law for dislocation creep:
\begin{equation}
	\eta_v = A^{-\frac{1}{n}} \left(\varepsilon^{II} \right)^{\frac{1}{n}-1} \exp \left( \frac{Q+pV}{nRT} \right)
\label{eq:Arrh-viscosity}
\end{equation}
where $A$, $n$ and $Q$ are material dependant parameters, $V$ is the activation volume of the material, $T$ is the temperature and $R$ is the universal gas constant. 

To assess which of the plastic ($\eta_p$) or viscous ($\eta_v$) viscosities is used to evaluate the stress, the minimum of the two is kept
\begin{equation}
	\eta = \min \left(\eta_p, \eta_v \right)   	
\end{equation}   

\subsection{Initial conditions} 
\label{sub:initial_conditions}

\begin{figure}[h!]
    \centering
    \includegraphics[scale=0.9]{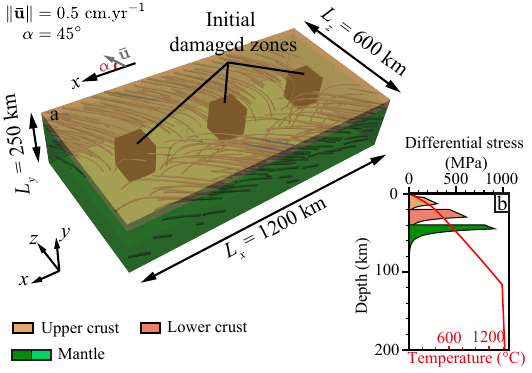}
    \caption{(a) 3D view of the modelled domain. 
    The 3 cubes show the initial geometry and location of the weak zones.
    The streamlines show the initial velocity field.
    (b) Yield stress envelope of the modelled lithosphere for a constant $\varepsilon^{II} = 10^{-14}$ s$^{-1}$ and a temperature distribution showed by the red curve.}
\end{figure}

The modelled domain $\Omega$ is a rectangular parallelepiped domain of dimensions $[O_x, L_x] \times [O_y, L_y] \times [O_z, L_z]$ discretized in $m_x \times m_y \times m_z$ $Q_2$ elements.
The material initial distribution consists of flat layers of different Earth's material.
The continental crust is divided into an upper crust extending from $ 0 \leqslant y < -20$ km, simulated with a quartz flow law \citep{Ranalli1997} and a lower crust extending from $-20 \leqslant y < -35$ km simulated with an anorthite flow law \citep{Rybacki2000}.
The mantle is also divided into two layers with the same olivine flow law \citep{Hirth2003} rheology. 
The lithosphere extends at coordinates $ -35 \leqslant y < -120 $ km and the asthenosphere at $ -120 \leqslant y \leqslant -250$ km.
\begin{table}
	\begin{center}
	\begin{tabular}{l l l}
		\hline
		Parameter & Units & Value\\
		\hline
		$[O_x, L_x]$ & km & $[0, 1200]$\\
		$[O_y, L_y]$ & km & $[-250, 0]$\\
		$[O_z, L_z]$ & km & $[0, 600]$\\
		$[m_x, m_y, m_z]$ & - & $[256, 64, 128]$\\
		\hline
	\end{tabular}
\end{center}
\caption{Model dimensions}
\label{tab:dim}
\end{table}
To initialise deformation and favour the formation of offset basins we define three areas in which a small random amount of plastic strain is set. 
This initial plastic strain will reduce the strength of the material due to plastic strain softening (Eq.~\ref{eq:softening}) and initialise deformation.

Moreover, the initial condition for the temperature field is computed using the steady-state heat equation 
\begin{equation}
	\nabla \cdot (k \nabla T) + H = 0
\label{eq:heat-steady}
\end{equation}
with the Dirichlet boundary conditions $T=0^{\circ}$C $\forall y = L_y$ and $T=1450^{\circ}$C $\forall y = O_y$. 
In addition, below the lithosphere, the heat transfer are known to be mainly advective due to active mantle convection leading to an adiabatic temperature gradient of approximately $0.5^{\circ}/$km.
To obtain such adiabatic gradient with the steady-state diffusive heat equation we enforce a very large thermal conductivity in the asthenosphere ($k=70$ W m$^{-1}$ k$^{-1}$).
However, for the time dependant solve of the conservation of energy (Eq.~\ref{eq:energy}) we use the realistic value given in Table~\ref{tab:param}.

\begin{table}
	\begin{center}
	\begin{tabular}{l p{2cm} p{2cm} p{2cm} p{3cm} p{3cm}}
		\hline
		Parameter & Units & Upper crust & Lower crust & Lithospheric mantle & Asthenospheric mantle\\
		\hline
		$A$ & MPa$^{-n}$.s$^{-1}$ & $6.7 \times 10^{-6}$ & $13.4637$ & $2.5 \times 10^{4}$ & $2.5 \times 10^{4}$\\
		$n$ & - & $2.4$ & $3$ & $3.5$ & $3.5$\\
		$Q$ & kJ.mol$^{-1}$ & $156$ & $345$ & $532$ & $532$\\
		$V$ & m$^3$.mol$^{-1}$ & $0$ & $3.8 \times 10^{-5}$ & $8 \times 10^{-6}$ & $8 \times 10^{-6}$\\
		$C_0$ & MPa & $300$ & $300$ & $300$ & $300$\\ 
		$C_{\infty}$ & MPa & $20$ & $20$ & $20$ & $20$\\
		$\epsilon_i$ & - & $0$ & $0$ & $0$ & $0$\\
		$\epsilon_e$ & - & $1$ & $1$ & $1$ & $1$\\
		$\beta$ & Pa$^{-1}$ & $10^{-11}$ & $10^{-11}$ & $10^{-11}$ & $10^{-11}$\\
		$\alpha$ & K$^{-1}$ & $3 \times 10^{-5}$ & $3 \times 10^{-5}$ & $3 \times 10^{-5}$ & $3 \times 10^{-5}$\\
		$k$ & W.m$^{-1}$.K$^{-1}$ & $2.7$ & $2.85$ & $3.3$ & $3.3$\\
		$H$ & $\mu$W.m$^{-3}$ & $1.5$ & $0.3$ & $0$ & $0$\\
		$\rho_0$ & kg.m$^{-3}$ & $2700$ & $2850$ & $3300$ & $3300$\\
		\hline
	\end{tabular}
\end{center}
\caption{Rheological and thermal parameters}
\label{tab:param}
\end{table}

\subsection{Boundary conditions} 
\label{sub:boundary_conditions}

\subsubsection{Dirichlet boundary conditions} 
\label{ssub:dirichlet_boundary_conditions}

To model an oblique rift we introduce the obliquity through the boundary conditions as it avoids using any free-slip condition and allows reaching higher obliquities \citep{Jourdon2021}.
Along faces of normal $\vec z$ we use Eq.~\eqref{eq:analytical-u-oblique} to impose the Dirichlet boundary conditions with $\|\vec{\bar{u}}\| = 0.5$ cm.y$^{-1}$.

On the face of normal $\vec y$ located at $y = L_y$ we impose a free surface boundary condition $\tens{\sigma}\nb{} = \vec 0$.

On the face of normal $\vec y$ located at $y = O_y$ we impose a flow balancing the inflow and outflow of all the other faces except for the top free surface.
To do so we first compute
\begin{equation}
	F = \sum \int_S \vec u \cdot \nb{} \, dS
\end{equation}
of all faces except for the top surface and then we set the bottom velocity as
\begin{equation}
	u_y = \frac{F}{L_x L_z}.
\end{equation}

In addition, in the model using only Dirichlet boundary conditions (Model D) we impose the linear velocity function described by Eq.~\eqref{eq:analytical-u-oblique} along faces of normal $\vec x$.

\subsubsection{Navier-slip boundary conditions} 
\label{ssub:navier_slip_boundary_conditions}
In the model using the generalised Navier-slip boundary conditions (Model GNS) we impose, along faces of normal $\vec x$, the following  boundary conditions
\begin{align}
	\vec u \cdot \nhat{} &= 0 &\forall x=\{0,L_x\}\\
	\tens{G}_S &= \tens{\mathcal{H}} \odot \left(\tens{\Lambda}^T \tens{\tau}_S \tens{\Lambda} \right) &\forall x=\{0,L_x\}
\end{align}
with 
\begin{equation}
	\begin{matrix}
		\that{} = 
		\begin{bmatrix}
			\bar{u}_{x_0}\\
			0\\
			\bar{u}_{z_0}
		\end{bmatrix}
		,&
		\nhat{} = \tens{R}_y \left(\frac{\pi}{2} \right) \that{},
		&
		\thati{} = \nhat{} \times \that{}
	\end{matrix}
\end{equation}
and 
\begin{equation}
	\tens{\mathcal{H}} = 
	\begin{bmatrix}
		0 & 1 & 1\\
		1 & 1 & 1\\
		1 & 1 & 0
	\end{bmatrix}
\end{equation}
and $\tens{\tau}_S$ is computed using a virtual strain rate tensor computed from Eq.~\eqref{eq:analytical-u-oblique} and described by Eq.~\eqref{eq:analytical-eps-oblique}. 
However, to obtain $\tens{\tau}_S$ we also require a viscosity which is neither constant nor linear.
Thus, we use the viscosity updated after each non-linear iteration.

\subsection{Results} 
\label{sub:results}
\begin{figure}[h!]
    \centering
    \includegraphics[scale=0.9]{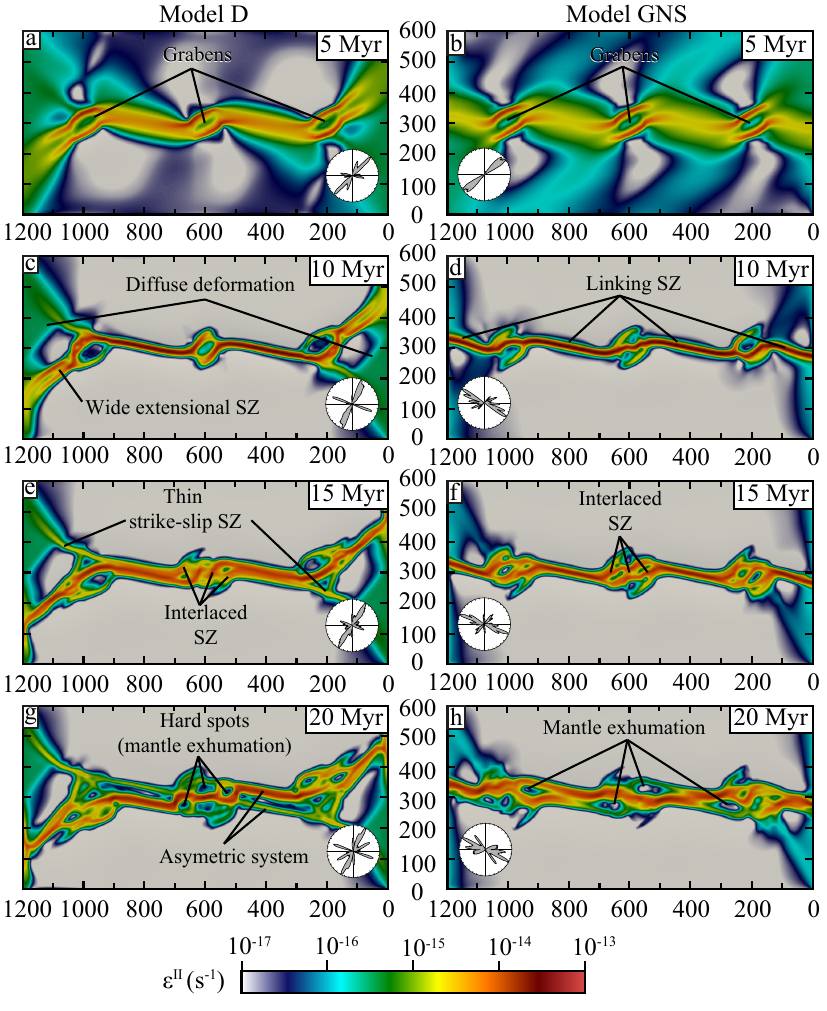}
    \caption{Map view at the surface of the model of $\varepsilon^{II}$ (Eq.~\ref{eq:eps-II}).
    The small circles show the orientation of the shear zones at the surface of the model in a stereoplot.
    Left: Model D.
    Right: Model GNS.
    SZ: shear zone}
    \label{fig:e2-models}
\end{figure}
\subsubsection{Model D} 
\label{ssub:model_d}
This model uses a linear velocity function along the faces of normal $\boldsymbol x$ promoting diffuse deformation near the boundaries of the model (Figure~\ref{fig:e2-models} left column).

During the first 10 Myr of evolution, the initial central weak zone located at $x = 600$ km mainly localises extensional deformation in favour of shear zones oriented N20 (Figure~\ref{fig:e2-models}a).
It is surrounded on both sides by two localising shear zones oriented $\sim$N110 (Figure~\ref{fig:e2-models}c).
Closer to the boundaries, the most distal initial weak zones located at $x = 200$ km and $x = 1000$ km are connecting to the domain boundaries with weakly localised shear zones oriented $\sim$N30-N50 forming a triangular shape (Figure~\ref{fig:e2-models}c).

From 10 Myr, the deformation in the shear zones joining the central weak zone ($x=600$ km) to the distal ones ($x=200$ km and $x=1000$ km) progressively forms two parallel shear zones accommodating strike-slip and extensional deformation (Figure~\ref{fig:e2-models}e, g).
It evolves into an asymmetric system in which the most localised shear zone thins the lithosphere and exhumes the mantle, marking the separation between an upper plate and a lower plate in which a thin strike-slip shear zone localises.

In the central part of the domain, the deformation evolves into a network of interlaced shear zones exhuming the mantle which generates hard spots around which the shear zones turn and localise (Figure~\ref{fig:e2-models}e, g).

Around the weak zones located closer to the boundaries of the domain, the deformation is partitioned between a thin strike-slip shear zone and a wider extensional shear zone forming an angle between each other of $\sim 50^{\circ}$ to $60^{\circ}$ (Figure~\ref{fig:e2-models}c, e).
The wide extensional shear zone remains oriented $\sim$N30 during the whole simulation, forming an angle of $\sim 75^{\circ}$ with direction of extension of the boundaries (Figure~\ref{fig:e2-models}c).
In contrast, the thin strike-slip shear zones developing between the boundaries and the initial distal weak zone are oriented from N110 near the weak zones to N130 near the boundaries forming an angle of less than $10^{\circ}$ with the extension direction (Figure~\ref{fig:e2-models}e).

\subsubsection{Model GNS} 
\label{ssub:model_gns}
\begin{figure}[h!]
    \centering
    \includegraphics[scale=0.9]{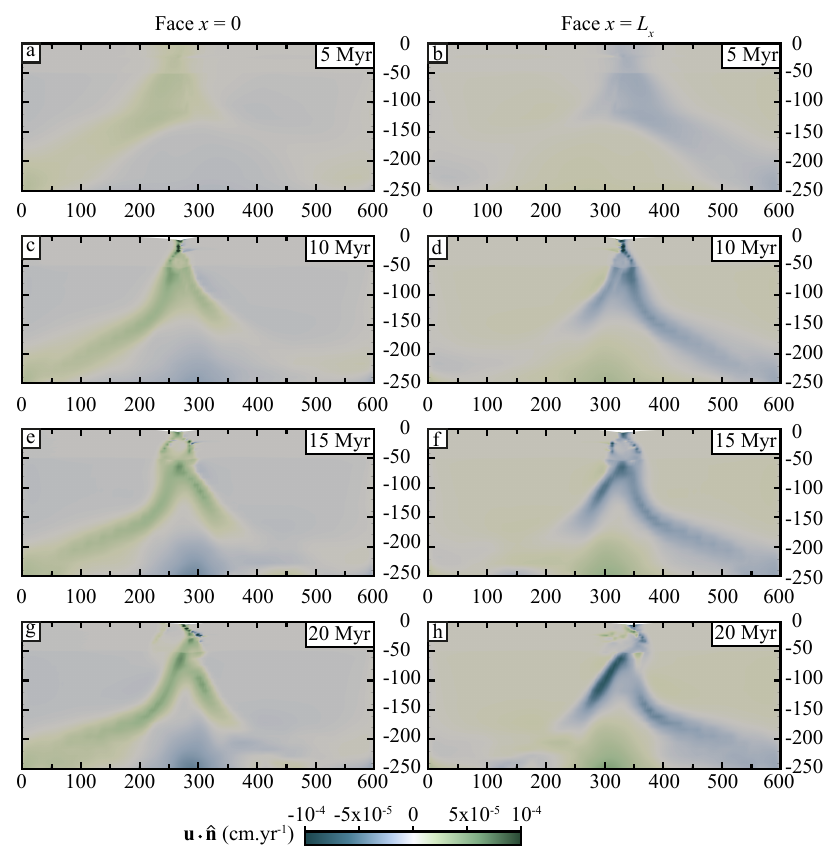}
    \caption{Evolution of $\vec u \cdot \nhat{}$ in cm.yr$^{-1}$ through time and space on face of normal $\vec x$.
    Non-linear residual stopping condition: $10^{-3}$. 
    Left column: face $x = O_x$. Right column: face $x = L_x$.}
    \label{fig:u.n-faces}
\end{figure}
This model uses the generalised Navier-slip boundary condition to impose an oblique extension along faces of normal $\boldsymbol x$ combined with Dirichlet boundary conditions along faces of normal $\boldsymbol z$.

During the first 10 Myr strain localises along extensional shear zones oriented between N40 and N60 forming grabens at the location of the three initial weak zones (Figure~\ref{fig:e2-models}b). 
At the edges of these grabens, deformation starts to localise along shear zones oriented N115 linking the central graben with the distal ones but also linking the most distal grabens with the domain boundaries (Figure~\ref{fig:e2-models}d).

From 10 Myr these shear zones start to propagate in the centre of each graben.
At 15 Myr in the grabens new shear zones form at their centre as interlaced extensional shear zones accommodating the thinning of the lithosphere and the exhumation of the mantle between them (Figure~\ref{fig:e2-models}f).

Finally, the shear zones between the grabens partition the deformation between strike-slip to transtensional segments and extensional segments leading to crustal break-up and exhumation of the mantle (Figure~\ref{fig:e2-models}h). 

\subsubsection{Behaviour of the solution along boundaries} 
\label{ssub:behviour_of_the_solution_along_boundaries}
\begin{figure}[h!]
    \centering
    \includegraphics[scale=0.9]{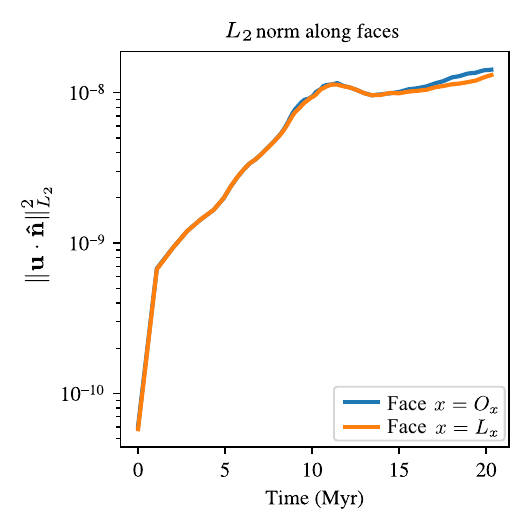}
    \caption{Evolution of $\|\vec u \cdot \nhat{}\|^2_{L_2}$ through time on faces of normal $\boldsymbol x$ for the model GNS.}
    \label{fig:l2norm-u.n}
\end{figure}

In Model GNS we weakly enforce that $\vec{u} \cdot \nhat{} = 0$ along the boundaries of normal $\boldsymbol x$. 
The solution shows that $|\vec{u} \cdot \nhat{}| < 10^{-4}$ cm.yr$^{-1}$ during the whole simulation (Figure~\ref{fig:u.n-faces}).
Moreover, Figure~\ref{fig:l2norm-u.n} shows that $\|\vec u \cdot \nhat{}\|^2_{L_2} \leqslant 10^{-8}$ at all time on both faces. 
Figure~\ref{fig:l2norm-u.n} also shows that it is more difficult to maintain very low values of $\vec u \cdot \nhat{}$ during strain localisation while remaining within acceptable order of magnitudes.


\subsubsection{Comparison between models} 
\label{ssub:comparison_between_models}
\begin{figure}[h!]
    \centering
    \includegraphics[scale=0.9]{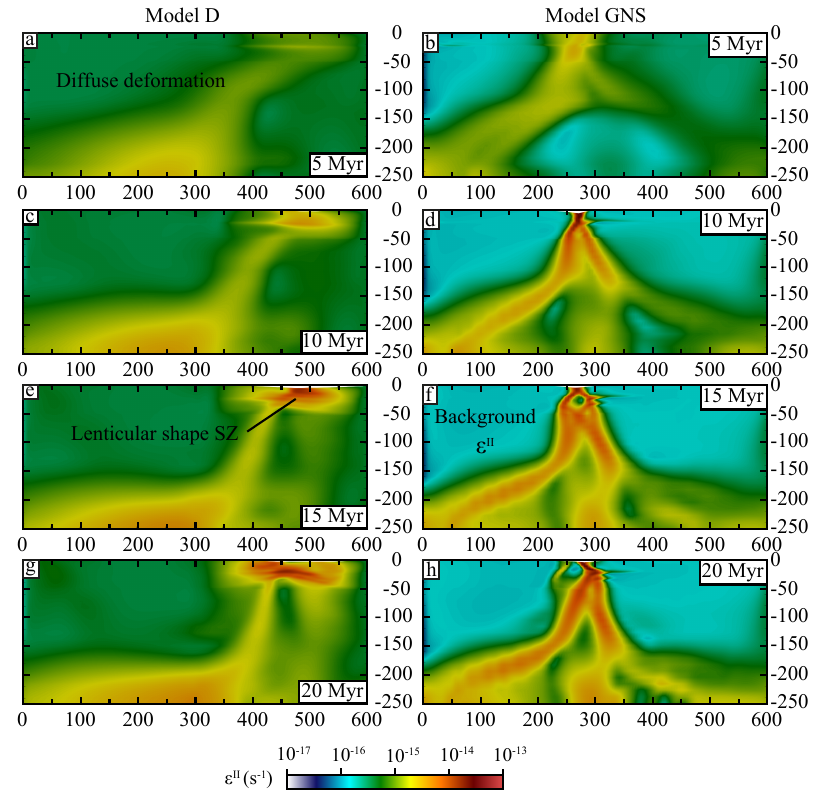}
    \caption{Active deformation (Eq.~\ref{eq:eps-II}) along the face $x = O_x$ of the model D (left) and model GNS (right).}
    \label{fig:e2-XMIN}
\end{figure}
The largest difference between models is located at and near the domain boundaries where either Dirichlet or generalised Navier-slip conditions are applied.
On the boundaries, Model D shows a diffuse deformation characteristic of a linear velocity distribution (Figure~\ref{fig:e2-XMIN}a).
In fact, because the Dirichlet boundary conditions we provide imposes the horizontal components of the velocity as linear functions, the only component that can actually generate strain localisation is the vertical one. 
Indeed, $\varepsilon^{II}$ on the faces of normal $\boldsymbol x$ shows that strain localises with lenticular shapes stretched in the $\boldsymbol z$ direction and shortened in the $\boldsymbol y$ direction (Figure~\ref{fig:e2-XMIN}e, g) which can be directly related to the linear variation of the velocity in the $\boldsymbol z$ and $\boldsymbol x$ direction and to the fact that the vertical component of the velocity remains unconstrained on these faces.

On the contrary, model GNS shows strain localisation patterns that are almost periodic i.e., the deformation between the initial weak zones is very similar to the deformation between the distal weak zones and the boundaries (Figure~\ref{fig:e2-models} right column).
This result would also be expected if the domain was larger and contained more equally spaced weak zones.
Moreover, the strain localisation and the shear zones formation on the boundaries of normal $\boldsymbol x$ evolve in agreement with the structures formed inside the domain and can migrate and rearrange themselves during the simulation (Figure~\ref{fig:e2-XMIN} right column).
However, the background low strain rate values on the boundaries remain higher than the low values inside the domain due to the stress boundary data introduced by the method with the $\tens{G}_S$ tensor. 

Finally, on the boundaries, the generalised Navier-slip boundary conditions allow localising strain in agreement with the system evolution while Dirichlet boundary conditions impose the solution once and for all.

Nevertheless, as showed on Figure~\ref{fig:e2-models} the differences between models are not only confined to the boundaries of the domain.
While the central initial weak zone area (around $x=600$ km) shows a close evolution between models, the linkage between grabens and the distal grabens themselves show a different evolution.
On the one hand, in Model D the evolution of distal grabens is different than the central one due to the boundary conditions influencing the shear zones orientation and the stress regime.
On the other hand, Model GNS shows a continuity and similarity in the strain pattern in the three grabens and the linking shear zones between them.

\section{Solver performance} 
\label{sec:solver_performance}
\begin{table}
	\begin{center}
	\begin{tabular}{m{2.5cm} m{2.5cm}|m{2.5cm}|m{2.5cm}|m{2.5cm}}
		\hline
		& 
		\multicolumn{2}{c|}{Generalised Navier-slip Model} & \multicolumn{2}{c}{Dirichlet Model}\\
		\hline
		 & Saddle point solve & $\tens{A}$ solve & Saddle point solve & $\tens{A}$ solve\\
		\hline
		$\#$ time steps & \multicolumn{2}{c|}{1,427}& \multicolumn{2}{c}{1,479}\\
		Total solves & 7,310 & 64,103 & 7,540 & 64,132\\
		Total its & 64,103 & 679,382 & 64,132 & 546,784\\
		min. its & 1 & 1 & 2 & 4\\
		max. its & 20 & 31 & 14 & 25\\
		Average its & 8.77 & 10.60 & 8.50 & 8.52\\
		std. dev. its & 4.19 & 2.65 & 2.60 & 2.93\\
		var. its & 17.58 & 7.02 & 6.75 & 8.58\\
		\hline
	\end{tabular}
\end{center}
\caption{Statistics of the saddle point and the viscous block $\tens{A}$ solves of Model GNS and Model D.}
\label{tab:solver-perf}
\end{table}

To solve the non-linear Stokes problem a Picard linearisation is applied to Eqs.~\eqref{eq:non-linear-residual} resulting in an approximate Jacobian matrix equivalent to the linear Stokes problem \citep{May2015}:
\begin{equation}
	\tens{J} = 
	\begin{bmatrix}
		\tens{A} & \tens{B}\\
		\tens{B^T} & \tens{0}
	\end{bmatrix}
\end{equation}
forming a saddle point problem.
Instead of being solved simultaneously for both velocity and pressure in a single process the system is first decomposed and solved for the viscous block $\tens{A}$ using the Krylov method FGMRES preconditioned with a geometric multigrid.

In table~\ref{tab:solver-perf} we compare the solver performance between the model using only Dirichlet boundary conditions and the model using the generalised Navier-slip boundary conditions.

In both models stopping condition of the residual for the saddle point problem convergence is based on a relative decrease of $10^{-4}$ between the first and the last iteration. 
Table~\ref{tab:solver-perf} shows that this residual decrease is, in average, achieved between 8 and 9 iterations for each saddle point solve for both models.

For the solve of the viscous block $\tens{A}$, the stopping condition for the residual is based on a relative decrease of $10^{-3}$ between the first and the last iteration.
The model using the generalised Navier-slip boundary conditions shows an average number of iteration between 10 and 11 to achieve this convergence while the model using only Dirichlet boundary conditions shows an average between 8 and 9 iterations.

While the evolution of the solution during the simulation is very different between the two models, their performance remains very close.
It shows that the use of the generalised Navier-slip boundary conditions strongly influences the result while poorly impacting the solver performance.

In addition, from experimental tests it appears that it is necessary to provide at least one component per stress vector in the stress tensor $\tens{\tau}_S$ (Eq.~\ref{eq:stress-data}) to be applied on the boundaries while using the generalised Navier-slip condition.


\section{Conclusion} 
\label{sec:conclusion}
In this work we introduced a new approach to apply Navier-slip boundary conditions in arbitrary directions based on Nitsche's method.
These boundary conditions require a direction in which the velocity is constrained as well as a stress field in the coordinate system defined by that arbitrary direction.
Imposing stress instead of velocity reduces the impact of the boundary conditions on the solution.
The magnitude of the velocity and its orientation are solved for in agreement with the stress data, the direction imposed and the evolution of the velocity field inside the domain.

Models results we provide in this study show that along the faces on which generalised Navier-slip boundary conditions are imposed, the velocity field behaves as if the domain was continuous instead of being bounded.
The methods allows applying oblique boundary conditions in regional geodynamic models without having to provide arbitrary velocity functions that strongly influence the evolution of the deformation in the model. 
Moreover, we show that using these boundary conditions does not affect the system stability and that classical Krylov methods preconditioned with multigrid algorithms can be used efficiently.



\bibliography{References}




\end{document}